\documentclass{article}
\usepackage{caption}
\usepackage{graphicx}
\usepackage{amsfonts}
\usepackage{amsmath}
\usepackage{amssymb}
\usepackage{mychicago}
\usepackage{genetics_manu_style}

\title{Adaptation in tunably rugged fitness landscapes: \\ The Rough Mount Fuji Model}  

\author{Johannes Neidhart\thanks{Institut f{\"u}r Theoretische Physik,
    Universit{\"at} zu K{\"o}ln, D-50937 K{\"o}ln, Germany}, Ivan G. Szendro$^\ast$ and Joachim Krug$^\ast$}

\renewcommand{\CorrespondingAddress}{Institut f{\"u}r Theoretische
  Physik, 
    Universit{\"at} zu K{\"o}ln \\
    Z{\"u}lpicher Str. 77, 
    D-50937 K{\"o}ln, Germany
    Phone +49 (0)221 470 2818\\
    \texttt{krug@thp.uni-koeln.de} \vfill}
\renewcommand{\RunningHead}{Rough Mount Fuji Model of fitness landscapes}
\renewcommand{\CorrespondingAuthor}{Joachim Krug}
\renewcommand{\KeyWords}{Fitness landscapes; epistasis; adaptive
  walks; experimental evolution}

\begin{document}

\maketitle

\begin{abstract}

Much of the current theory of adaptation is based on Gillespie's
mutational landscape model (MLM), which assumes that the fitness values of
genotypes linked by single mutational steps are independent random
variables. On the other hand, a growing body of empirical evidence shows that
real fitness landscapes, while possessing a considerable amount of
ruggedness, are smoother than predicted by the MLM. In the present
article we propose and analyse a simple fitness landscape model with
tunable ruggedness based on the Rough Mount Fuji (RMF) model
originally introduced by Aita et al. [Biopolymers \textbf{54}:64-79 (2000)] in the context of protein
evolution. We provide a comprehensive collection of results pertaining
to the topographical structure of RMF landscapes, including explicit
formulae for the expected number of local fitness maxima, the location
of the global peak, and the fitness correlation function. The
statistics of single and multiple adaptive steps on the RMF landscape
are explored mainly through simulations, and the results are compared
to the known behavior in the MLM model. Finally, we show that the RMF
model can explain the large number of second-step mutations observed
on a highly-fit first step backgound in a recent evolution experiment 
with a microvirid bacteriophage [Miller et al., Genetics
\textbf{187}:185-202 (2011)].  

\end{abstract}


The genetic adaptation of an asexual population to a novel environment is governed by
the number and fitness effects of available beneficial mutations,
their epistatic interactions, and the rate at which they are
supplied \cite{Sniegowski2010}. Despite the inherent complexity of
this process, recent theoretical work has identified several robust
statistical patterns of adaptive evolution \cite{Orr2005a,Orr2005b}.
Most of these predictions were derived in the
framework of Gillespie's mutational landscape model (MLM), which is based on
three key assumptions \cite{gillespie,gillespie2,gillbook,orr1}. 
First, selection is strong enough to prevent the
fixation of deleterious mutations and mutation is sufficiently weak
such that mutations emerge and fix one at a time (the strong
selection/weak mutation or SSWM regime). Second, wildtype fitness is
high, which allows one to describe the statistics of beneficial
mutations using extreme value theory (EVT). Third, the fitness values of
new mutants are uncorrelated with the fitness of the ancestor from
which they arise. This last assumption implies that the fitness
landscape underlying the adaptive process is maximally rugged with
many local maxima and minima \cite{kaufflev,kauffman,Jain2007}, a
limiting situation that is often referred to as the House of Cards
(HoC) landscape \cite{kingman2}. Thus, the MLM is concerned with a population 
evolving in a HoC landscape under SSWM dynamics, starting 
from an initial state of high fitness.  

The validity of the SSWM assumption depends primarily on the
population size $N$. Denoting the mutation rate by $u$ and the typical
selection strength by $s$, the criterion for the SSWM regime reads
$Nu \ll 1 \ll Ns$, which can always be satisfied by a suitable choice
of $N$ provided $u \ll s$, as is usually the case. On the other
hand, whether or not the other two assumptions underlying the MLM are
realistic is an empirical question that has been addressed in a number
of experimental studies of microbial evolution. Investigations
aimed at determining the distribution of effect sizes of benefical mutations 
have generally found support for the EVT hypothesis \cite{Orr2003,joyceorr,Orr2010},
and examples for all three EVT universality classes have been reported
in the literature
\cite{Rokyta2005,Kassen2006,rokyta2,MacLean2009,Schenk2012,Foll2014,Bank2014}. 
At the same time, however, it has become increasingly clear that the extreme
assumption of uncorrelated fitness values between genotypes connected
by single mutational steps cannot be upheld in the face of empirical
evidence. 

Indications for the presence of correlations in real fitness
landscapes derive from two types of experimental studies. In one
approach, a subset of the fitness landscape is explicitly
generated by constructing genotypes containing all combinations 
of a small group of mutations chosen for either individual or 
collective effects, and measuring their fitness or some
proxy thereof \cite{Weinreich2006,Lozovsky2009,jaspjoa,Schenk2013,Szendro2012,Weinreich2013,deVisser2014}. 
Although the topographic properties of the resulting 
landscapes vary over a broad range, in most cases they display a
degree of ruggedness that is intermediate between a smooth, additive
landscape and the maximally rugged landscape assumed by the MLM
\cite{Szendro2012,deVisser2014}. In a second approach, properties of the underlying
landscape are inferred from the observed dynamics of adaptation as
manifested, for example, in the trajectories of fitness increase \cite{Kryazhimskiy2009}
or the number of substitutions in an adaptive walk \cite{Gifford2011,Schoustra2012}. Of
particular interest in the present context is the recent study of
\citeN{miller} on the microvirid bacteriophage ID11, where the MLM was tested by comparing the
distribution of beneficial mutations from the wildtype to the
corresponding distribution after one step of adaptation. According to
the MLM, the two distributions should be identical up to a rescaling,
but this hypothesis was clearly refuted by the experiment. 

The observation that most empirical fitness landscapes display an
intermediate degree of ruggedness implies that there is a need for
simple, mathematically tractable landscape models in which the
ruggedness can be tuned. A frequently used model with tunable ruggedness
is Kauffman's ``NK''-landscape where each of $L$ binary loci
interacts randomly with $K$ other loci, and the interaction degree $K$
serves to interpolate between the additive limit $K=0$ and the HoC
limit $K = L-1$ 
\cite{Kauffman1989,kauffman,Ohta1997,Welch2005,Aita2008,Oestman2012,Franke2012}.
In the original definition of the NK-model the letter `N' stands for the sequence length which we 
denote by $L$ in the present paper. While this model has
been shown to be capable of describing various features of empirical
fitness landscapes \cite{Hayashi2006,Rowe2010,jaspjoa}, its mathematical complexity is such
that even rather elementary properties -- for example, the mean number
of local fitness maxima \cite{Evans2002,Durrett2003,Limic2004} -- are difficult to derive in closed form (but
see \citeN{Perelson1995}, \citeN{Orr2006} and \citeN{Schmiegelt2014} for a variant of the
NK-model that is simpler to analyze).  

In the present article we therefore propose the Rough Mount Fuji (RMF)
model as an alternative description of fitness landscapes with tunable
ruggedness. The model is a simplified
version of the RMF fitness landscape originally introduced by \citeN{aita} 
in the context of protein evolution. In essence,
the RMF model superimposes an additive fitness landscape and an
uncorrelated random HoC landscape, and the ruggedness
is tuned by changing the ratio of the additive selection coefficient
to the standard deviation of the random fitness component
\cite{Aita2000,jaspjoa,Szendro2012}. Below we derive simple, explicit formulae for
various quantitative measures of the RMF topography such as the number and
location of local fitness maxima and fitness correlations. Moreover,
assuming SSWM conditions, 
we show how the adaptation of a population on the RMF landscape can be efficiently
simulated for realistic numbers of loci by locally generating the
mutational neighborhood of the current genotype along the adaptive
trajectory. Finally, as an example for the application of the RMF
model to empirical fitness landscapes, we estimate the parameters 
of the fitness landscape of the microvirid bacteriophage ID11 studied by
\citeN{miller} by matching the number of secondary beneficial
mutations available after one adaptive step (the \textit{number of
exceedances}) predicted by the model to the experimentally
observed value. 


\section{Model}

\subsection{Definition.}
Following a common practice in the description of empirical fitness
landscapes, we represent genotypes by binary sequences $\sigma =
(\sigma_1, \sigma_2, ....,\sigma_L)$ of fixed length
$L$, composed of elements taken from the set $\{0,1\}$
with $\sigma_i = 1$ ($\sigma_i = 0$) if a mutation is present (absent) 
at the $i$'th locus. The set all binary sequences of length $L$ is
known as the Hamming space. It is endowed with a natural distance
measure, the \textit{Hamming distance} defined as 
\begin{equation}
\label{d}
D(\sigma,\sigma') = \sum_{j=1}^L (\sigma_j - \sigma'_j)^2
\end{equation}
which simply counts number the loci at which $\sigma$ and $\sigma'$
differ. It is convenient to introduce the antipodal (or reversal)
sequence $\overline{\sigma}$ of $\sigma$ through $\overline{\sigma}_i =
1 - \sigma_i$. A sequence and its antipode are maximally distant from
each other, $D(\sigma,\overline{\sigma}) = L$. 

To introduce the Rough Mount Fuji model we first choose a reference
sequence $\sigma^\ast$ which represents the state of maximal fitness
of the additive part of the fitness landscape. The fitness $F(\sigma)$ of genotype
$\sigma$ is then defined through 
\begin{equation}
\label{RMF}
F(\sigma) =  -c D(\sigma, \sigma^\ast)+\eta(\sigma),
\end{equation}
where $c > 0$ is a constant parameter and the $\eta$'s are $2^L$
independent and identically distributed (i.i.d.) random
variables. Equation \ref{RMF} describes an average decrease of
fitness with increasing distance from $\sigma^\ast$ by an amount of
$c$ per mutational step, superimposed by a random fitness
variation. For $c = 0$ the RMF model reduces to an uncorrelated HoC
landscape, while for large $c$ it becomes essentially additive, as the
random fitness component $\eta$ is then negligible compared to the mean
fitness gradient. The competition between the additive and random contributions
is governed by the parameter 
\begin{equation}
\label{theta}
\theta = \frac{c}{\sqrt{\mathrm{Var}(\eta)}}
\end{equation}
defined as the ratio between the additive selection coefficient $c$ and the standard
deviation of the random fitness component $\eta$ \cite{jaspjoa}. With increasing
$\theta$ the landscape becomes less rugged.

It is important to note that the RMF landscape is anisotropic, in the
sense that the mutational neighborhood of a sequence $\sigma$ depends
on its distance from $\sigma^\ast$. To be specific, we define the 
neighborhood $\nu(\sigma)$ of $\sigma$ as the set
$\nu(\sigma) = \{\sigma'|D(\sigma, \sigma')=1\}\cup
\{\sigma\}$. Denoting the distance of $\sigma$ from the reference
sequence by $d \equiv D(\sigma,\sigma^\ast) > 0$, the set $\nu(\sigma)$ is split
into three parts:
\begin{enumerate}
    \item the \textit{downhill neighborhood} that consists of the
      $L-d$ sequences at distance $d+1$ from $\sigma^\ast$, which have
      an expected  fitness
    disadvantage of $c$ compared to $\sigma$,
    \item $\sigma$ itself, and 
    \item the \textit{uphill neighborhood} that consists of the $d$
      sequences at distance $d-1$ from $\sigma^\ast$ which have an expected
      fitness advantage of
    $c$ compared to $\sigma$.
\end{enumerate}
This decomposition implies that, in contrast to the MLM,  the fitness values of the mutational
neighbors of $\sigma$ are not i.i.d. random variables. We will see in
the following how this leads to new properties of the fitness
landscape and of the adaptive process on that landscape.

\subsection{Fitness distribution and extreme value theory.}
To complete the definition of the RMF model we need to specify the
statistics of the random fitness component $\eta$ in terms of its
probability distribution function $P(x) \equiv \mathbb{P}(\eta < x)$ and
the corresponding probability density $p(x) = \frac{d}{dx} P(x)$.
Following the approach developed previously for the MLM, we invoke extreme value theory
(EVT) to classify the fitness distribution according to its
tail behavior. The underlying reasoning is that viable organisms must have high
fitness in absolute terms, which implies that the beneficial mutations that drive adaptation
reside in the tail of the (usually unknown) distribution of fitness effect of all possible
mutations \cite{gillespie,gillespie2,orr1,Orr2003,Orr2010}. The theorems of EVT then show that,
irrespective of the detailed form of the full distribution, the tail shape has to fall into
one of three classes \cite{haanferreira}:
\begin{itemize}
    \item The \textit{Gumbel class} containing all distributions with
      unbounded support and a density vanishing faster than a power law, for example, the
exponential, normal and gamma distributions; 
    \item the \textit{Fr\'echet class} containing all distributions
      with unbounded support and a density vanishing as a power law, and
    \item the \textit{Weibull class} containing all distributions with a truncated upper tail.
\end{itemize}
The three classes are conveniently represented through the Generalized
Pareto Distribution (GPD) defined by the distribution function \cite{Pickands1975,Beisel2007,joyceorr}
\begin{equation}
\label{GPD}
P_\kappa(x)=1-(1+\kappa x)^{-\frac 1 \kappa}  
\end{equation}
with the extreme value index $\kappa$. For $\kappa > 0$ the support of
$P_\kappa$ is $[0,\infty)$ and the distribution belongs to the
Fr\'echet class, while for $\kappa < 0$ the support is $[0,
-\frac{1}{\kappa}]$ and the distribution belongs to the Weibull
class. For $\kappa \to 0$ the GPD reduces to an exponential which
is a representative of the Gumbel class.

In the initial formulation of the EVT approach it was argued that the Gumbel
class is the most likely case to be realized biologically, and empirical support for
this hypothesis was found in several studies 
\cite{Rokyta2005,Kassen2006,MacLean2009,Orr2010}. However, subsequently systems
showing truncated (Weibull-type) fitness distributions were discovered \cite{rokyta2},
and recently several examples for heavy-tailed (Fr\'echet-type) fitness distributions
have been reported \cite{Schenk2012,Foll2014,Bank2014}. A truncated fitness distribution arises 
naturally in the adaptation towards a single fitness peak, as assumed in Fisher's geometric 
model, and analysis of this model shows that the distribution becomes Gumbel-like as the dimensionality of the underlying phenotype increases \cite{Orr2006b,Martin2008}. A similar
mechanistic interpretation is not known for distributions falling into the Fr\'echet class,
but the available empirical evidence suggests that such heavy-tailed distributions may generally
be associated with situations of strong selection pressure, e.g., in the adaptation
of pathogens to new drugs \cite{Schenk2012,Foll2014,Bank2014}.

We will see below that 
several properties of the RMF fitness landscape take on a particularly simple
form when the random fitness component is chosen from a particular
representative of the Gumbel class, the 
\textit{Gumbel distribution} defined by 
\begin{equation}
\label{eq:Gumbel}
P_G(x) = e^{-e^{-x}}, \;\;\; p_G(x) = e^{-x-e^{-x}}.
\end{equation}
This distribution arises in EVT as the limit law of the maximum of
i.i.d. random variables drawn from a distribution in the Gumbel
class \cite{haanferreira}. 
It approaches an exponential for large positive values and vanishes
very rapidly (as the exponential of an exponential) for negative values.
The key property of interest here is the behavior of $P_G$ under
shifts, 
\begin{equation}
\label{eq:Pshift}
P_G(x + a) = P_G(x)^{e^{-a}}.
\end{equation}
For completeness we note that the mean of the Gumbel density
$p_G(x)$ is the Euler-Mascheroni constant $\gamma \approx 0.5772156649...$
and its variance is $\frac{\pi^2}{6} \approx 1.644934067...$. 
We emphasize that our use of this distribution in the following sections 
is motivated primarily by its mathematical convenience.

\subsection{Comparison to empirical fitness landscapes.}
Several recent studies using the RMF model to quantify properties of empirical fitness landscapes
provide some guidance as to what range of model parameters can be expected in applications
to biological data. \citeN{jaspjoa} and \citeN{Szendro2013a} analyzed an 8-locus fitness
landscape composed of individually deleterious mutations in the filamentous fungus 
\textit{Aspergillus niger}. Based on the statistical properties of selectively accessible mutational
pathways and a direct fit to the data, respectively, and assuming a normal distribution for the
random fitness component $\eta$, they obtained consistent estimates $\theta \approx 0.25$ and 
$\theta \approx 0.21$ for the parameter defined in Equation \ref{theta}. In a metaanalysis of 10 different
fitness data sets the \textit{A. niger} landscape was found to be among the more rugged empirical
landscapes, indicating that these estimates for $\theta$ are at the lower end of the range of 
biologically relevant values \cite{Szendro2012}. However, we will see below that the properties
of the RMF landscape are strongly affected by the EVT index $\kappa$, which implies that 
both $\theta$ and $\kappa$ are generally required to characterize an empirical
data set (see THE NUMBER OF EXCEEDANCES for an example). 
 
\section{Structure of the fitness landscape}
In this section we present results concerning the main structural
features of the RMF fitness landscape, in particular its local maxima
and fitness correlations. 

\subsection{Fitness Maxima.}
Local fitness maxima play a key role in adaptation, as they present obstacles to an evolving population, and the number of maxima is a
commonly used measure of landscape ruggedness. In the HoC model all genotype fitness values are
independent and statistically equivalent. The
probability that a given sequence is a local fitness maximum is then simply
$\frac{1}{L+1}$, since each of the $L+1$ sequences in the neighborhood
are equally likely to have the largest fitness, and the expected
number of maxima is $\frac{2^L}{L+1}$ \cite{kaufflev,kauffman}. 
These expressions for the maximally rugged HoC model serve as a
benchmark for the corresponding results for the RMF model that will be
presented in the following. Detailed derivations are given in APPENDIX A.

\subsubsection{Density of local maxima.}
\begin{figure}

    \begin{minipage}{.49\textwidth}
    \includegraphics[width=\textwidth]{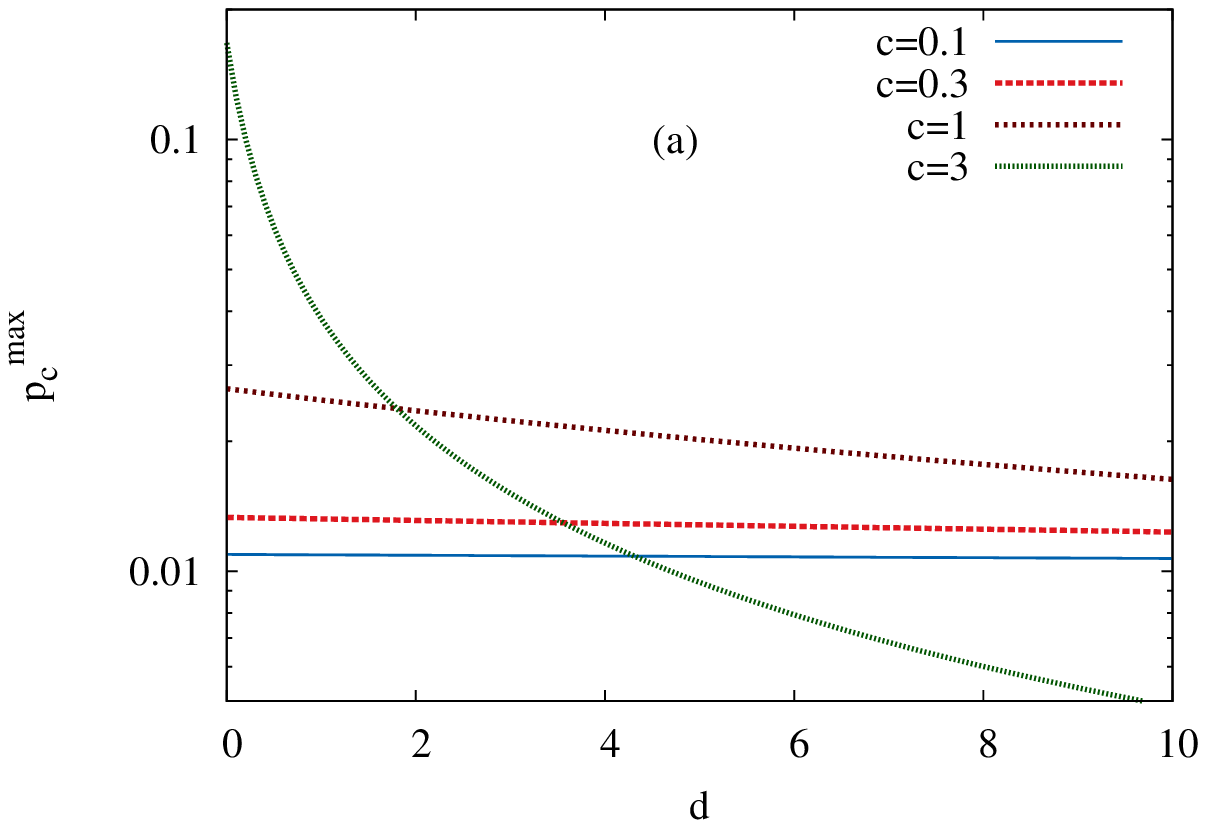}    
    \end{minipage}
    \hspace{.5cm}
    \begin{minipage}{.49\textwidth}
    \includegraphics[width=\textwidth]{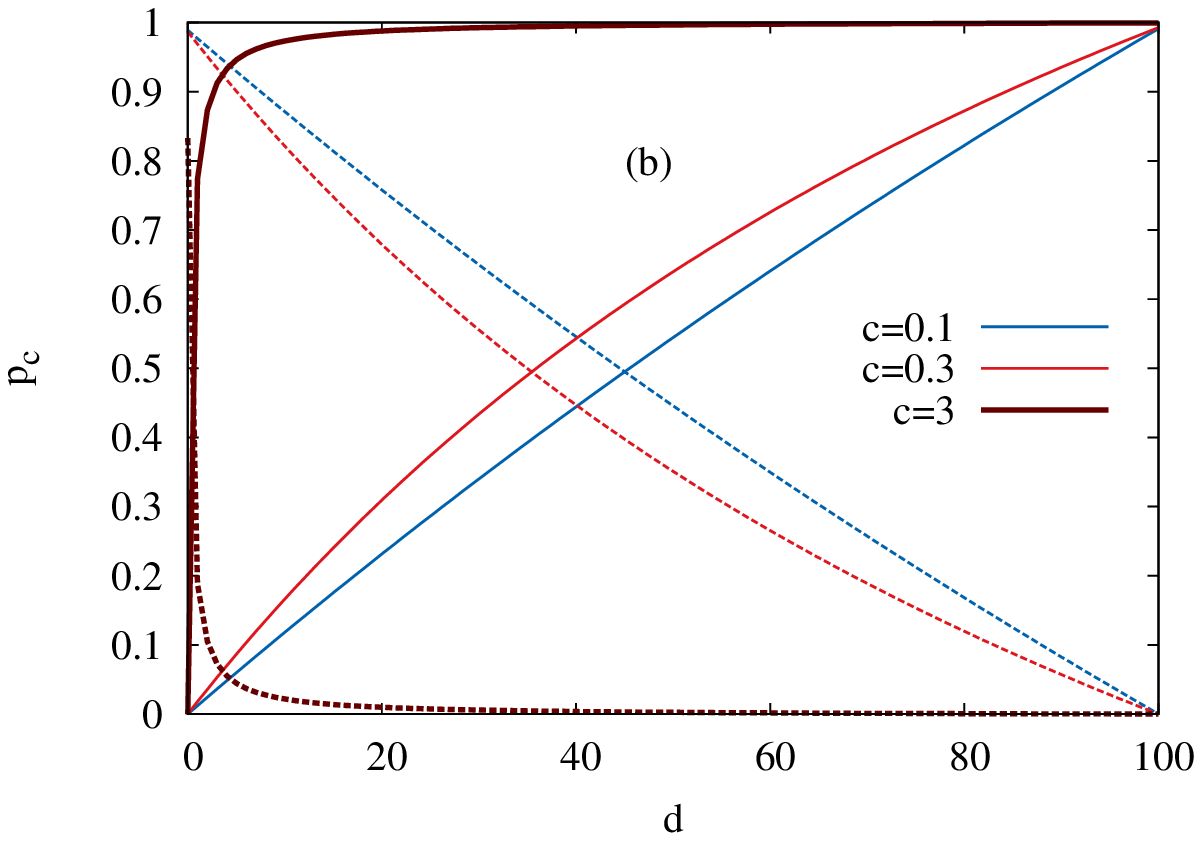}
    \end{minipage}
\caption{\label{fig:pc}(a) The probability that a given sequence is a local fitness
  maximum is shown as a function of the distance $d$ to the  reference
  sequence for several values of $c$ and $L=100$. The density of maxima is enhanced
in the vicinity of the reference sequence, and this effect becomes more pronounced
with increasing $c$. (b) The probability
  that the neighboring sequence of largest fitness is in the uphill
  (solid lines) or downhill (dashed lines) direction is shown as a
  function of $d$ for different values of $c$ and $L=100$. This quantity determines
the local direction of a greedy adaptive walk. In both
  panels the random fitness component is Gumbel distributed.}
\end{figure}

In the RMF model the probability $p^\text{max}_c(d)$ that a given
genotype is a local fitness maximum depends on its distance $d$ to the
reference sequence. When the random component is Gumbel distributed
this quantity can be computed exactly, with the result
\begin{equation}
    p^\text{max}_c(d) = \frac{1}{1+d e^c + (L-d)e^{-c}}.\label{eq:pmaxgum}
\end{equation}
The limits of this expression for small and large $c$ 
\begin{align}
\label{eq:pmax-limits}
     p^\text{max}_c(d) &=\begin{cases}
                            \frac{1}{L+1} & c\to 0\\
                            \delta_{d,0},&c\to \infty
                         \end{cases}
\end{align}        
correspond to the HoC model and the additive landscape with a single
maximum at $d=0$, respectively. Here the Kronecker symbol $\delta_{x,y}$ is defined by
\begin{align}
\label{eq:Kronecker}
     \delta_{x,y} &=\begin{cases}
                            1  & x=y\\
                            0 & \textrm{else}.
                         \end{cases}
\end{align}    
The behavior of \eqref{eq:pmaxgum} for intermediate values of $c$ is
illustrated in Figure \ref{fig:pc} (a).
 

It is also of interest to consider the probability that the
neighboring genotype of largest fitness for a sequence at distance $d$ from
the reference sequence is located in the uphill or downhill part of
the neighborhood, denoted by $p_c^\text{up}(d)$ and
$p_c^\text{down}(d)$, respectively. These probabilities determine the
fate of a `greedy' adaptive walk that chooses the neighboring sequence
of highest fitness at each step \cite{orr2}. They 
can be explicitly evaluated for the Gumbel distribution, with the result
\begin{equation} \label{eq:pupdown_Gumbel}
    p_c^\text{up}(d) =\frac{d}{d+e^{-c}+e^{-2c}(L-d)}, \;\;\;\;\;\;
    p_c^\text{down}(d)=\frac{L-d}{L-d+e^c+e^{2c}d}.
\end{equation}
Note that $p_c^\text{up} + p_c^\text{down} + p_c^\text{max} = 1$ and
$p_c^\text{up} = d e^c p_c^\text{max}$, $p_c^\text{down} = (L-d)
e^{-c} p_c^\text{max}$.
In the absence of a fitness gradient ($c=0$) the greedy walker is more 
likely to go uphill (downhill) for $d > L/2$ ($d < L/2$) because of
the greater availability of neighboring sequences in that
direction. For $c > 0$ the crossing point where $p_c^\text{up} =
p_c^\text{down}$ moves towards the reference sequence and is generally
located at $d=\frac{L}{1+e^{2c}}$, see Figure \ref{fig:pc}b).

\subsubsection{Total number of maxima.}

\begin{figure}
    \begin{minipage}{.49\textwidth}
    \includegraphics[width=\textwidth]{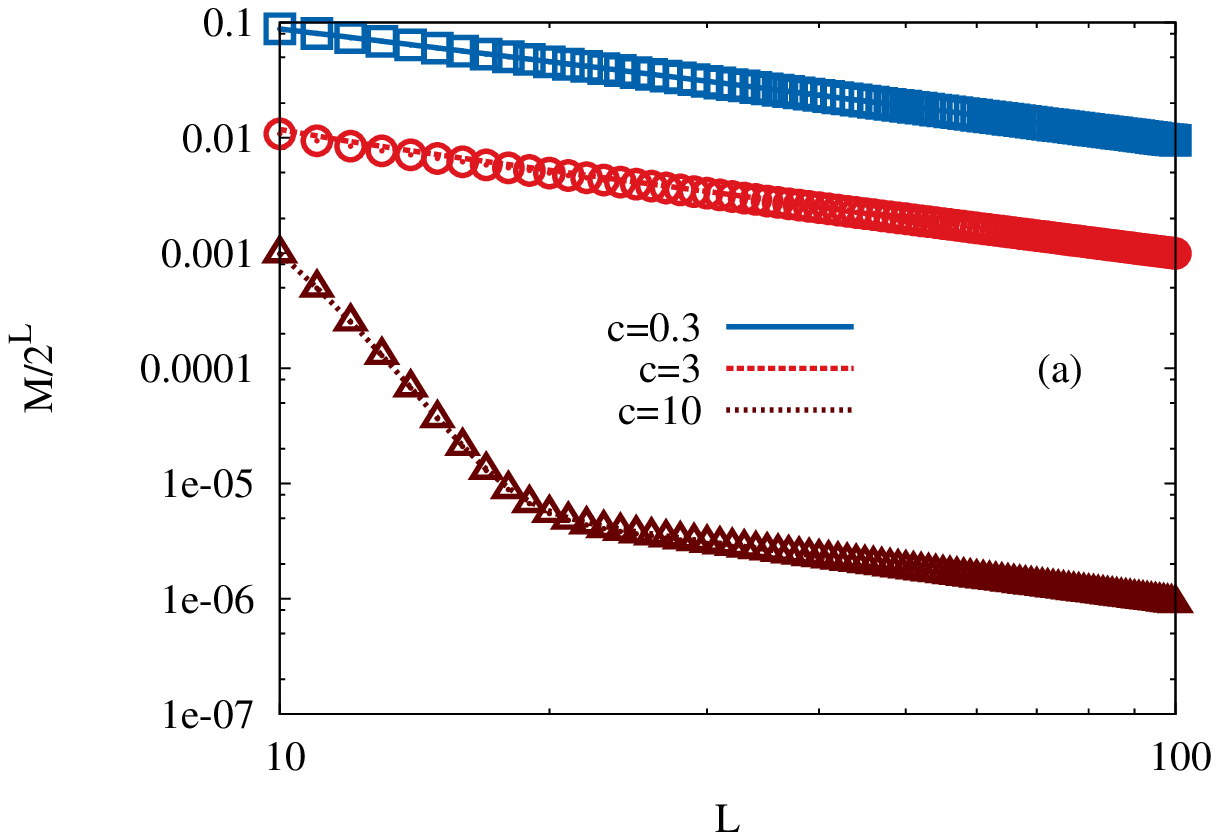}
    \end{minipage}
    \hspace{.5cm}
    \begin{minipage}{.49\textwidth}
    \includegraphics[width=\textwidth]{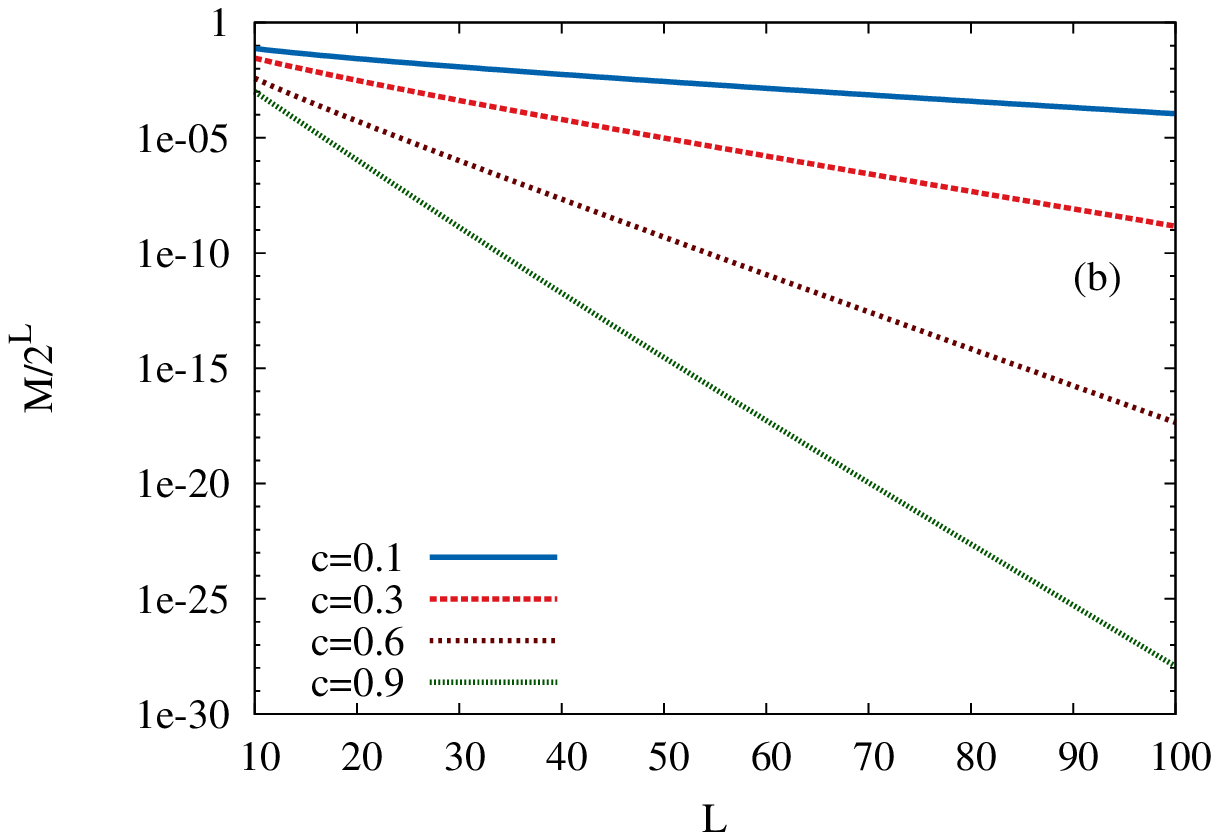}
    \end{minipage}
\caption{\label{fig:gum2} 
(a) The density of local fitness maxima ${\cal{M}}/2^L$ in a
      Gumbel distributed RMF landscape is shown as a function of the
      number of loci $L$. Symbols correspond to the large $L$
      approximation in Equation \ref{eq:Mapprox} 
    and lines to the exact expression in Equation \ref{eq:nomsum_Gumbel}. The double logarithmic
scales illustrate that the ratio ${\cal{M}}/2^L$ decays algebraically as $1/L$ for large 
$L$. (b) The same quantity for an RMF landscape with uniformaly distributed random component,
as given by the exact expressions in Equations \ref{M:uniform1} and \ref{M:uniform2}. 
Semi-logarithmic scales are used to illustrate the exponential decay of the ratio ${\cal{M}}/2^L$.}
\end{figure}

To determine the expected number of local fitness maxima $\mathcal M$ in
the entire landscape, the distance dependent probability $p_c^\mathrm{max}(d)$ has to be averaged over
$d$ with the appropriate weights giving the number of genotypes at
distance $d$,  
\begin{equation}
\label{eq:nomsum}
\mathcal M =\sum_{d=0}^{L}\binom{L}{d} p^\text{max}_c(d).
\end{equation}
Using the exact expression in Equation \ref{eq:pmaxgum} for the Gumbel distribution, the
sum in Equation \ref{eq:nomsum} can be expressed in terms of a hypergeometric
function, see Equation \ref{eq:nomsum_Gumbel}. To obtain a more tractable expression we note that
for large $L$ the binomial weights in
Equation \ref{eq:nomsum} become sharply peaked around $d= \frac{L}{2}$, such that 
 \begin{equation}
    \mathcal M \; \stackrel{L \to \infty}{\approx} \;  2^L
    p_c^\text{max}(L/2) = \frac{2^L}{L\cosh(c) +1}\label{eq:largeL}.
    \end{equation}
For fixed $L$ and large $c$ the expression in Equation \ref{eq:largeL} violates the obvious bound
$\mathcal M \geq 1$. A simple modification that cures this deficiency is
\begin{equation}
\label{eq:Mapprox}
\mathcal M_\text{approx} = 1 + \frac{2^L}{L\cosh(c) +1},
\end{equation}
which provides a very good approximation to the exact number of maxima
over the entire range of parameters, see Figure \ref{fig:gum2} (a). 
Interestingly, apart from a multiplicative constant the large $L$
asymptotics of the number of maxima is the same as for the HoC-model,
$\mathcal M \sim \frac{2^L}{L}$. This is in contrast to the corresponding
results for Kauffman's NK-model, where (depending on how the
number of interacting loci $K$ is scaled with the sequence length $L$)
different exponential and algebraic dependencies of the number of
maxima on $L$ can be found \cite{Perelson1995,Evans2002,Durrett2003,Limic2004,Schmiegelt2014}.

An exact expression for the expected number of maxima is derived in APPENDIX A for 
exponentially distributed randomness. In that case the asymptotic behavior
for large $L$ is seen to be identical to that in Equation \ref{eq:largeL}, and in fact this
behavior arises whenever the tail of the distribution is exponential
(see Figure \ref{fig:compnom} (a)). For tails heavier
than exponential it is shown that the asymptotics is exactly that of the HoC model, 
${\cal{M}} \to \frac{2^L}{L}$ independent of $c$, which implies, remarkably, that the 
mean fitness gradient has no effect on the number of maxima when the genotype dimension
is sufficiently large. This result applies in particular to distributions in the 
Fr\'echet class of EVT (Figure \ref{fig:compnom} (b)). For Gumbel-class distributions with tails thinner than exponential
the number of maxima is significantly reduced for any value of $c$, and the ratio
of $\cal{M}$ to the HoC value $2^L/L$ vanishes sub-exponentially in $L$ 
(see Equation \ref{M:genGum3}). 

The effect of the mean fitness gradient on the number of local maxima is most pronounced
for distributions with bounded support belonging to the Weibull class of EVT. 
An exact expression for the total 
number of maxima is derived in APPENDIX A for the case of a uniform fitness distribution,
a simple representative of this class, and the result is illustrated in 
Figure \ref{fig:gum2} (b). To leading order the number of maxima is proportional to $(2-c)^L/L$,
which interpolates smoothly between the HoC result for $c=0$ and the additive limit ${\cal{M}}= 1$ 
attained at $c=1$; for $c > 1$ the increase in fitness gained in one step towards the reference
sequence exceeds the support of the distribution of the random component, and the landscape becomes
strictly monotonic in $d$. For other distributions in the Weibull class with support on the
the unit interval the behavior is similar, ${\cal{M}} \sim (2-c^{-\frac{1}{\kappa}})^L$
to leading order (Equation \ref{M:genWeib2}). The behavior of the number of maxima for distributions with bounded support is thus reminiscent of the 
NK-model at fixed $K$, where it has been shown that ${\cal{M}} \sim \lambda_K^L$ with a 
$K$-dependent constant with $1 < \lambda_K < 2$ for $0 < K < L-1$
\cite{Durrett2003,Limic2004,Schmiegelt2014}. 

\subsubsection{Location of the global maximum.} We next ask where the
global maximum is located. For large $c$ it will be found close to the
reference sequence $\sigma^\ast$, while in the HoC limit $c=0$ it is
equally probable to be located anywhere. Since most sequences lie near
the distance $L/2$ from the reference sequence, also the global
maximum is then most likely at $d = \frac{L}{2}$. 
In general, the probability for the global maximum to lie at Hamming
distance $d$ from the reference sequence is given by
\begin{equation}
 P_\mathrm{max}(d)=\binom{L}{d}\tilde{P}_\mathrm{max}(d),
\end{equation}
where $\tilde{P}_\mathrm{max}(d)$ is the probability for some specific
genotype at distance $d$ to be the globally fittest state and the
binomial coefficient accounts for the multiplicity of states at
distance $d$. In APPENDIX A it is shown that for the Gumbel
distribution
\begin{equation}
\label{global_dist:Gumbel}
 \tilde{P}_\mathrm{max}(d) = \frac{e^{-cd}}{(1 + e^{-c})^L}.
\end{equation}
With this quantity at hand we proceed to calculate the mean distance of the global maximum from $\sigma^\ast$,
\begin{equation}
 \mathbb{E}(d) =\frac{L\mathrm{e}^{-c}}{1+\mathrm{e}^{-c}},
\end{equation}
which interpolates smoothly between the two limiting cases discussed
above, and the corresponding variance
\begin{equation}
\mathrm{Var}(d) =\frac{L\mathrm{e}^{-c}}{(1+\mathrm{e}^{-c})^2}.
\end{equation}
These calculations could be extended to include the
positions of sub-optimal local maxima and thus to address the possible
clustering of maxima discussed previously in the context of the 
NK-model \cite{kauffman}, but we do not pursue this question here. 
For general distributions an expansion for small $c$ 
shows that the shift in the position of the maximum from the HoC
value $L/2$ is of order $c L 2^{-\kappa L}$, where $\kappa$ where $\kappa$ is the extreme value
index defined in Equation \ref{GPD}. For distributions in the
Weibull class ($\kappa < 0$) this implies that minute values of $c
\sim 2^{\kappa L}$ suffice to bring the global maximum close to the reference sequence with high probability.

\subsection{Fitness correlations.}
In addition to the number of local maxima, a commonly used measure for
fitness landscape ruggedness is the decay of fitness correlations
\cite{Weinberger1990,Stadler1999}. Here we consider the correlation
function defined by  
\begin{equation}
\label{eq:C(r)}
 C(r)=\frac{\langle(F(\sigma)-\langle
   F(\sigma)\rangle)(F(\sigma')-\langle F(\sigma') \rangle)\rangle_r}{\langle (F(\sigma)-\langle F(\sigma)\rangle)^2\rangle_r}
\end{equation}
where angular brackets denote an average over sequence space as well as over the realizations of the random fitness component
$\eta$ in Equation \ref{RMF}, and $\langle \cdot \rangle_r$ denotes an
average over pairs of sequences $\sigma,\sigma'$ with $D(\sigma,\sigma') = r$.
The normalization of the expression in Equation \ref{eq:C(r)} ensures that $C(0) = 1$. 
The derivation in APPENDIX B shows that the correlation function depends
on the underlying random fitness distribution only through its variance,
and it is given by the expression
\begin{equation}
\label{eq:CRMF}
 C_\text{RMF}(r)=\frac{\frac{\theta^2}{4}(L-2r)+\delta_{r,0}}{\frac{\theta^2L}{4}+1}.
\end{equation}
The correlation function is a superposition of a peak at $r=0$, which
originates from the uncorrelated HoC component in Equation \ref{RMF}, and a
linearly decaying piece that reflects the global fitness gradient. 
It is instructive to compare this result to the correlation function
for the NK-model, which reads \cite{Campos2002,Campos2003}
\begin{equation}
\label{eq:CNK}
 C_\text{NK}(r)=\binom{L}{r}^{-1} \binom{L-K-1}{r}.
\end{equation}
This displays a linear decay, $C_\text{NK}(r) = 1 - \frac{r}{L}$, in
the non-epistatic limit $K=0$. However, in contrast to Equation \ref{eq:CNK}
which is non-negative for any $r$, Equation \ref{eq:CRMF} becomes negative
for $r > \frac{L}{2}$ (see also \citeN{Neidhart2013} for further
discussion of the relation between the two models). 

\section{Adaptation on the RMF landscape}

In the previous section we studied properties that depend purely on the topography of the 
landscape. Now we will shift the focus to the implications of 
the landscape structure for the evolutionary dynamics. More specifically, we will consider the dynamics in the SSWM limit. Here, only 
one genotype is populated at any time. If a beneficial mutation
occurs, it is either fixed in the entire population or the mutant goes
extinct before another mutation arises. Hence, the population behaves as a
single entity that performs an `adaptive walk' (AW) on the fitness
landscape \cite{gillespie}.

The adaptive walk is a sequence of single adaptive steps. In each
step, the population moves from the currently
populated sequence to a neighboring one with a transition probability
given by the fixation probability normalized by the fixation
probabilities of all other available beneficial mutants. Following
\citeN{gillespie} and \citeN{orr1} we consider the rank-ordered fitness
values $F_j$ in the current mutational neighborhood, where the rank of
the resident genotype is $i$ and beneficial mutants have ranks $j <
i$. Since the fixation probabilty of a beneficial mutation in the SSWM
regime is proportional to its selection coefficient, the  
probability for a transition from the current
$i$-th fittest genotype to the $j$-th fittest mutant is
\begin{align}
    P_{i\to j} = \frac{F_j-F_i}{\sum_{k=1}^i (F_k-F_i)} \label{eq:transprob}.
\end{align}
Based on this expression a number of results have been obtained for
the adaptation dynamics on the uncorrelated HoC landscape \cite{orr1,rokyta,joyceorr}. 
In the following we ask how these results are modified by the fitness
gradient in the RMF model.

\subsection{A single step of adaptation.}

\begin{figure*}
    \begin{minipage}{.49\textwidth}
    \includegraphics[width=\textwidth]{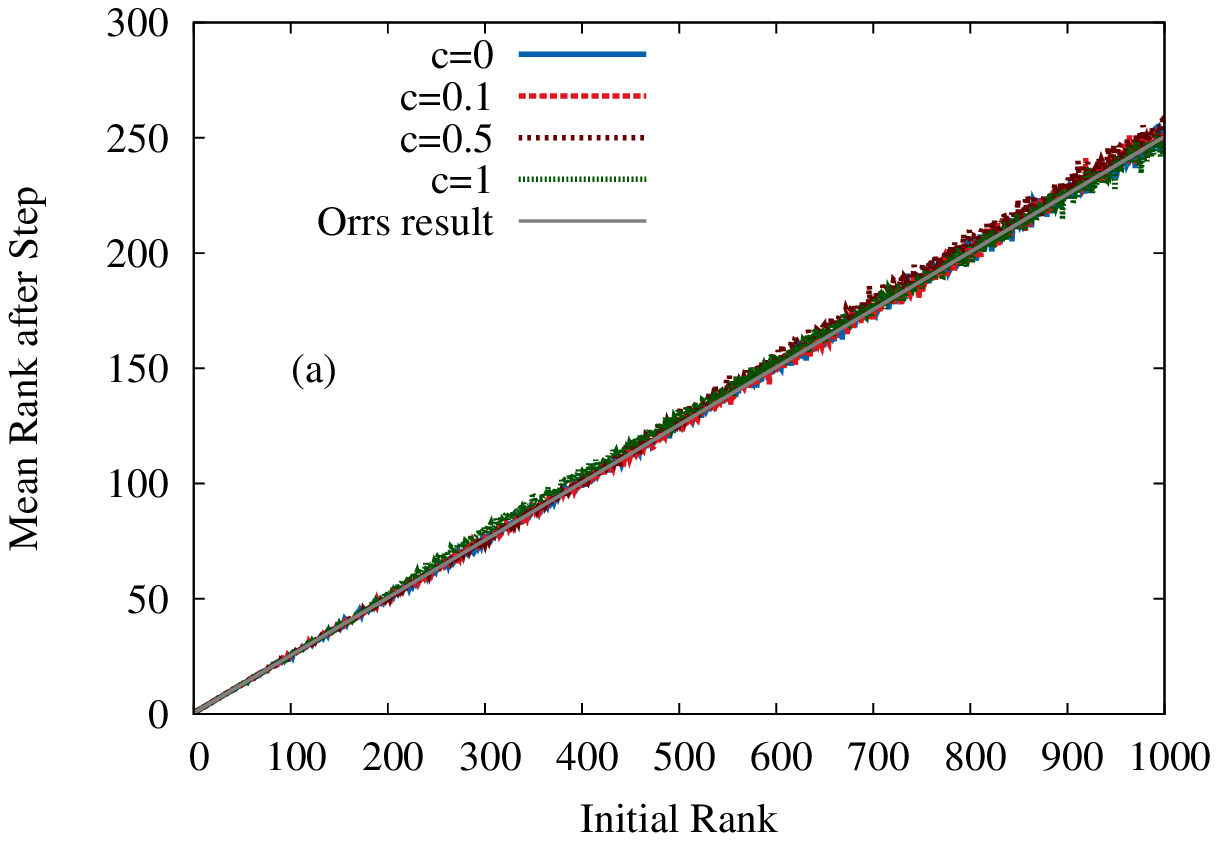}
    \end{minipage}
    \hspace{.5cm}
    \begin{minipage}{.49\textwidth}
    \includegraphics[width=\textwidth]{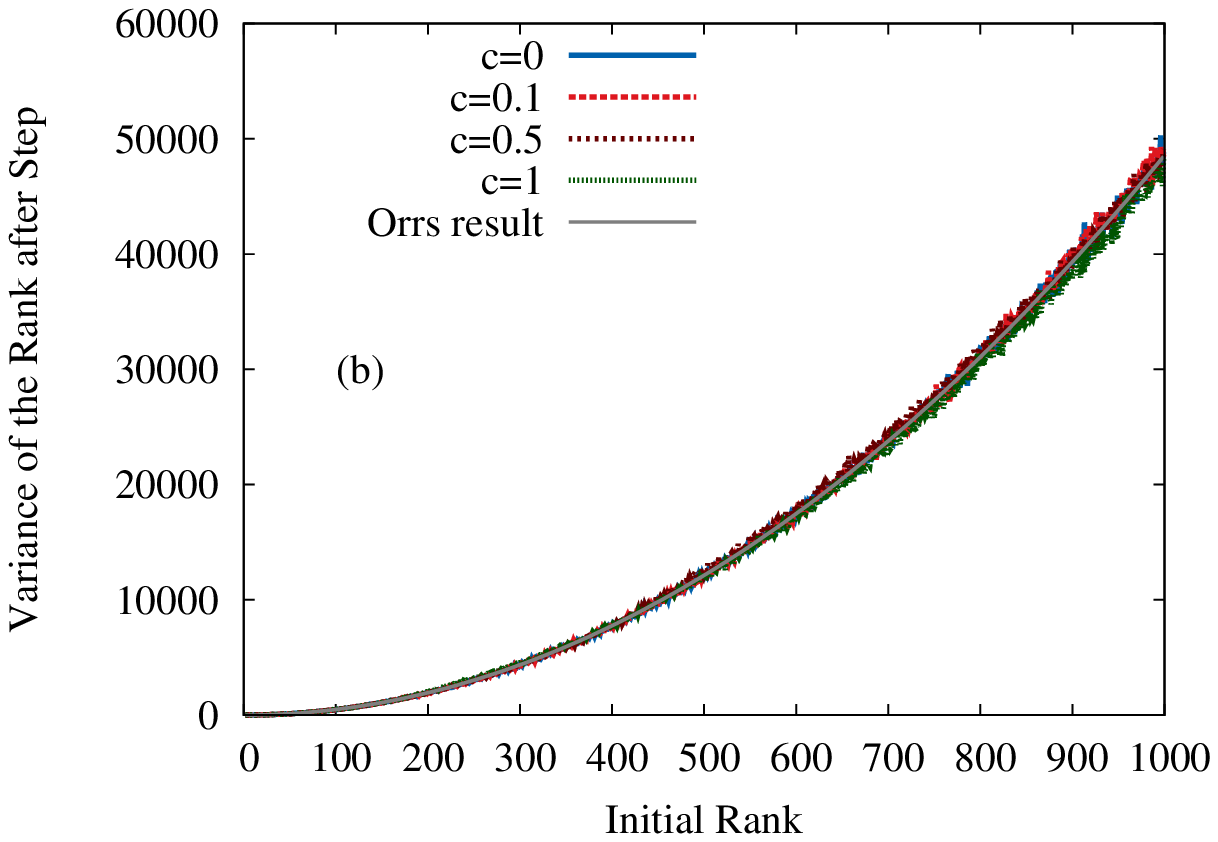}
    \end{minipage}
    \caption{(a) Mean and (b) variance of the fitness rank after an adaptive step in the exponentially
    distributed RMF landscape is shown as a function of the initial
    rank. Simulation results for different values of $c$ are 
    compared to the analytical expressions in Equation \ref{eq:orrrank} for the HoC
    landscape ($c=0$). The good agreement indicates that the properties of single adaptive
steps are only weakly affected by the fitness gradient. Here $L=1000$ and $d=50$.}
    \label{fig:rankchange}
\end{figure*}

\begin{figure*}
    \begin{minipage}{.49\textwidth}
    \includegraphics[width=\textwidth]{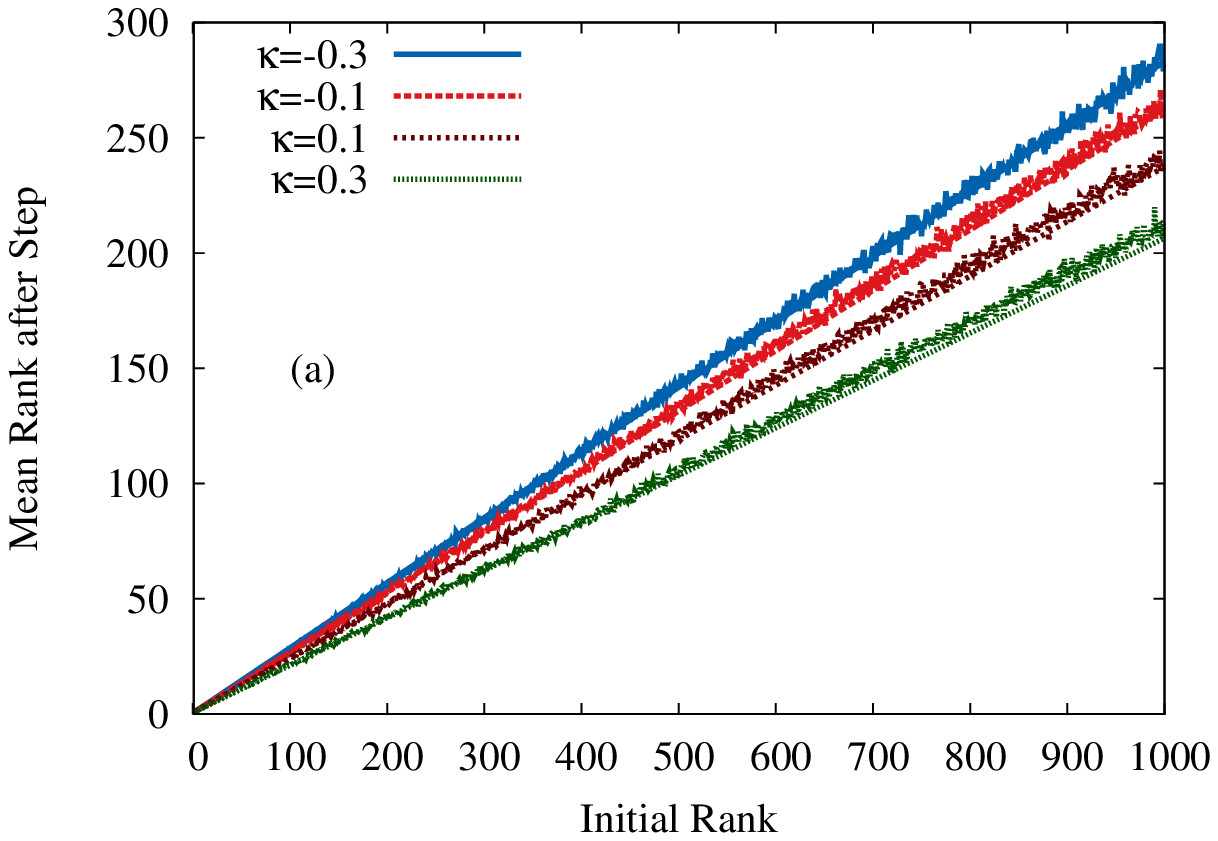}
    \end{minipage}
    \hspace{.5cm}
    \begin{minipage}{.49\textwidth}
    \includegraphics[width=\textwidth]{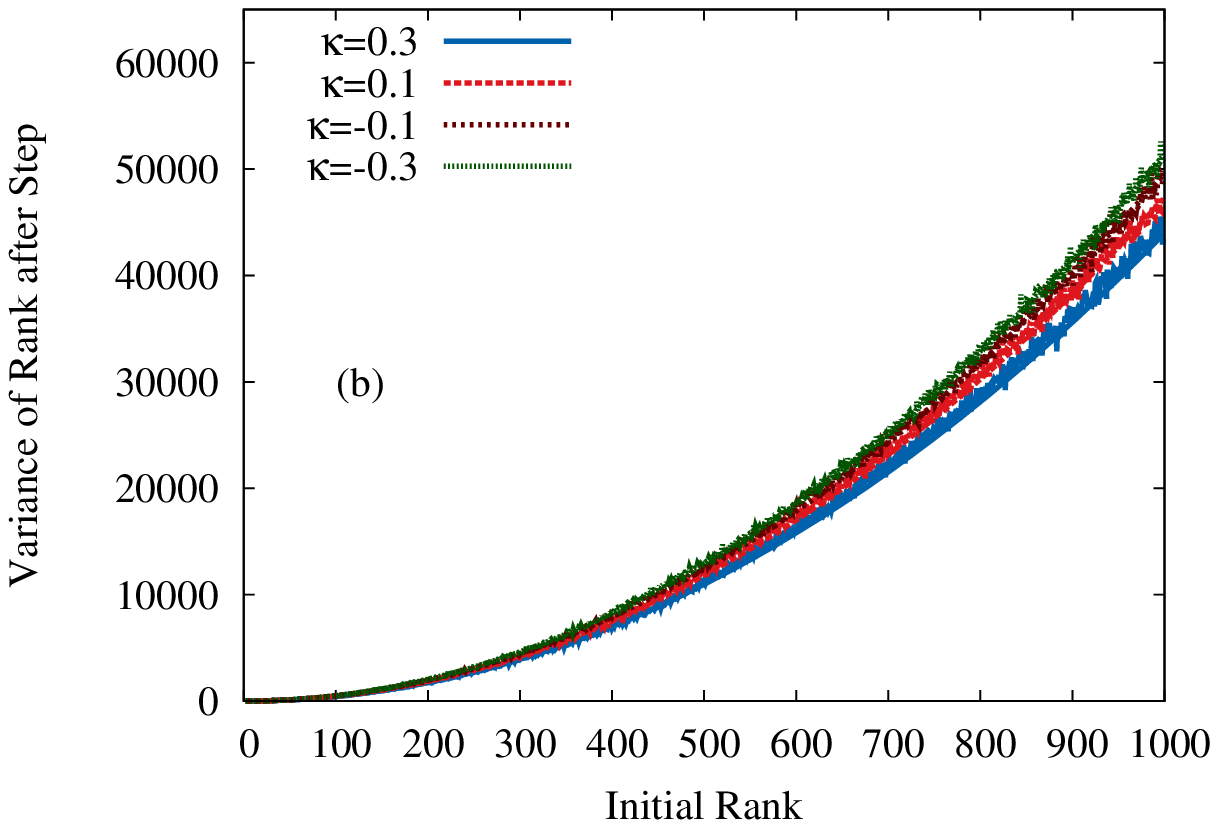}
    \end{minipage}
    \caption{(a) Mean and (b) variance of the fitness rank after an adaptive step in the GPD distributed RMF
    landscape is shown as a function of the initial rank. Simulation
    results for different values of $\kappa$ and $c=0.5$ are 
    compared to the analytical expressions in Equation \ref{eq:joycerank} for the HoC
    landscape, again showing close agreement. Here $L=1000$ and $d=50$.}
    \label{fig:gpdrankchange}
\end{figure*}

Before we turn to the full adaptive walk we consider a single step of
adaptation, specifically the change in the rank of the resident genotype during an adaptive step. 
Using the transition probability in Equation \ref{eq:transprob}, \citeN{orr1} calculated the expectation value and the
variance of the rank $j$ of the next populated sequence after an
adaptive step starting from the genotype with rank $i$. For HoC
landscapes with fitness values drawn from a Gumbel class
distribution he obtained
\begin{align}
    \mathbb{E}(j) = \frac{i+2}{4}, \phantom{aaaaaaaaa} \text{Var}(j)=\frac{(i-2)(7i+6)}{144}.\label{eq:orrrank}
\end{align}
These results were subsequently generalized to the other extreme value
classes by \citeN{joyceorr}, who found, for EVT index $\kappa < \frac{1}{2}$, 
\begin{align}
    \mathbb{E}(j)&=1+ \frac{i-2}{2}\left (\frac{1-\kappa}{2-\kappa}\right),\nonumber\\
    \text{Var}(j)&=\frac{(1-\kappa)(i-2)[(\kappa^2-4\kappa
    +7)i+6(1-\kappa)]}{12(3-\kappa)(2-\kappa)^2}.\label{eq:joycerank}
\end{align}
For $\kappa \to \frac{1}{2}$ the approximation used to derive the
expression for $\text{Var}(j)$ breaks down because the fitness
distribution ceases to have a second moment; similarly the expression
for $\mathbb{E}(j)$ breaks down for $\kappa \to 1$. For further
discussion of this case of extremely heavy-tailed fitness
distributions we refer to \citeN{Schenk2012}.

To see to what extent these results are modified in the RMF landscape
we refer to APPENDIX C, where it is shown that the distribution 
of fitness values in a mutational neighborhood of the RMF model with exponential
randomness approximately remains a simple exponential that is shifted
by a constant amount depending on the distance $d$ to the reference
sequence as well as on the model parameters $c$ and $L$.  
This implies that the statistics of the fitness \textit{spacings}
$F_{k-1} - F_k$ are approximately independent of $c$ and $d$ and take the
same form as in the HoC model with exponential fitness distribution.
In fact the exponential nature of the fitness spacings in this case 
is guaranteed by a general theorem of order statistics \cite{Smid1975}.  
Since the transition probability in Equation \ref{eq:transprob} can be written
in terms of these spacings,
it seems reasonable to conjecture that the properties of single adaptive steps  
in the RMF model are well approximated by the HoC results at least for
moderate value of the fitness gradient $c$. 

To test this conjecture, simulations consisting of the following steps
were carried out: (i) create a random RMF neighborhood, (ii)
determine the current rank of the initial genotype, (iii) carry out
an adaptive step according to Equation \ref{eq:transprob} and (iv) determine the
new rank in the old neighborhood.  
A comparison of the RMF simulation results to Orr's analytical formulae
in Equation \ref{eq:orrrank} is shown in Figure \ref{fig:rankchange}. Obviously,
the HoC expressions are in good
agreement with the simulation data even in cases where the slope
$c$ of the RMF landscape is comparable to the variance $v$ of the
random fitness contributions (for the exponential distribution
considered here $v = 1$). 

For other distributions the approximation in APPENDIX C is not
directly applicable, but the results
shown in Figure \ref{fig:gpdrankchange} indicate that the generalized
HoC formulae in Equation \ref{eq:joycerank} provide a good approximation to the 
RMF model for a range of EVT indices around $\kappa = 0$. Thus we
conclude, somewhat unexpectedly, that the statistical properties of
single adaptive steps are not strongly affected by the fitness
gradient in the RMF model. 

\subsection{Adaptive Walks.}
\begin{figure}
\begin{minipage}{\textwidth}
\begin{center}
    \includegraphics[width=.6\textwidth]{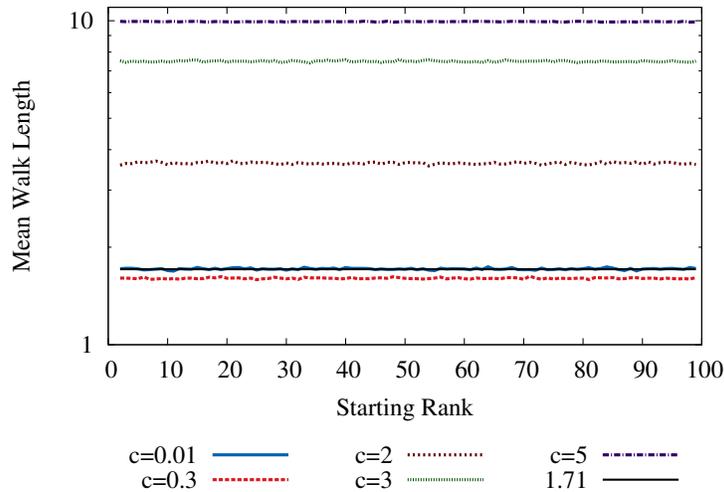}
\end{center}
    \end{minipage}
\caption{Mean length of greedy adaptive walks in an exponentially-distributed RMF
  landscape with $L=100$, $d=10$. The walk length is independent of
  starting rank for all $c$.}
\label{fig:gaw}
\end{figure}


In the context of AW's, a property of interest is the
\textit{mean walk length} $\ell$, which is the average number of steps performed
until the process reaches a local fitness maximum and terminates. The mean walk length 
in the HoC landscape has been analyzed using various approaches \cite{gillespie,orr1,flyvlaut,krugneid,jainsee,Jain2011,Seetharaman2013}. 
Because of the lack of isotropy in the RMF landscapes, analytical
results for the walk length are much harder to derive for this model, 
and the results described in the following were therefore obtained from simulations. 

The simulation algorithm is analogous to that described above for the first step of adaptation, but now a new
neighborhood is created after each adaptive step and the procedure is repeated until a local maximum has been found,
that is, until the current genotype has rank 1 in its new neighborhood. Since the distribution of the fitness
values in the new neighborhood depends on the distance to the reference genotype, the direction of the adaptive
steps (uphill or downhill) has to be kept track of. Creating the neighborhoods `on the fly' during the adaptive
walk implies that the memory of previously visited genotypes is lost beyond the second step. However, the error associated
with this approximation is expected to be negligible for large $L$ \cite{flyvlaut,krugneid} and it is the only
feasible approach for simulating walks on landscapes with thousands of loci. 

On HoC landscapes the mean walk length is determined primarily by the starting rank $r$, that is,
the rank that the first populated state has in its initial neighborhood, and was shown in previous work to be   
proportional to $\log(r)$ for $r \ll L$, see below for further discussion.
On the other hand for a smooth landscape with only one maximum, the walk length
equals of course the distance $d$ of the starting point to this maximum. Due to its anisotropic structure, on the RMF
landscape one expects the walk length to depend on both initial rank and initial distance to the reference sequence $\sigma^\ast$. 
Specifically, for small $c$ (in the sense $\theta \ll 1$), 
the mean adaptive walk length should increase logarithmically in the starting rank and be approximately independent of the initial
Hamming distance $d$ from $\sigma^\ast$, while for $\theta \gg 1$, 
it should increase linearly in $d$ and be approximately independent of the starting rank, 
since with high probability only a single maximum exists in the fitness landscape at $\sigma^\ast$ or close to it. Analysis of a simplified version of the problem where the walks are assumed to start from the
antipode $\overline{\sigma^\ast}$ of the reference sequence shows that the two regimes are separated
by a sharp transition under certain conditions \cite{Park2014}. 

Before discussing the process governed by the transition probability
in Equation \ref{eq:transprob}, we briefly consider the
simpler case of a \textit{greedy} adaptive walk in which the transition occurs deterministically to the available genotype of maximal
fitness in each step. On HoC landscapes, the mean walk length
for this process is known to be asymptotically constant for large $r$ and $L$, approaching the universal limit 
$\ell = e-1 \approx 1.72$ \cite{orr2}. For the RMF model, simulations displayed in Figure \ref{fig:gaw} show  
that for very small $c$, the walk length is still on average equal to $\ell=e-1$. For larger $c$, the
mean walk length first decreases slightly (see curve corresponding to $c=0.3$) and then increases rapidly, until $\ell =d$,
the limit expected for a smooth landscape. 
For all values of  $c$, the average walk length remains independent of the starting rank.

\begin{figure}
	\begin{minipage}{.49\textwidth}
		\includegraphics[width=\textwidth]{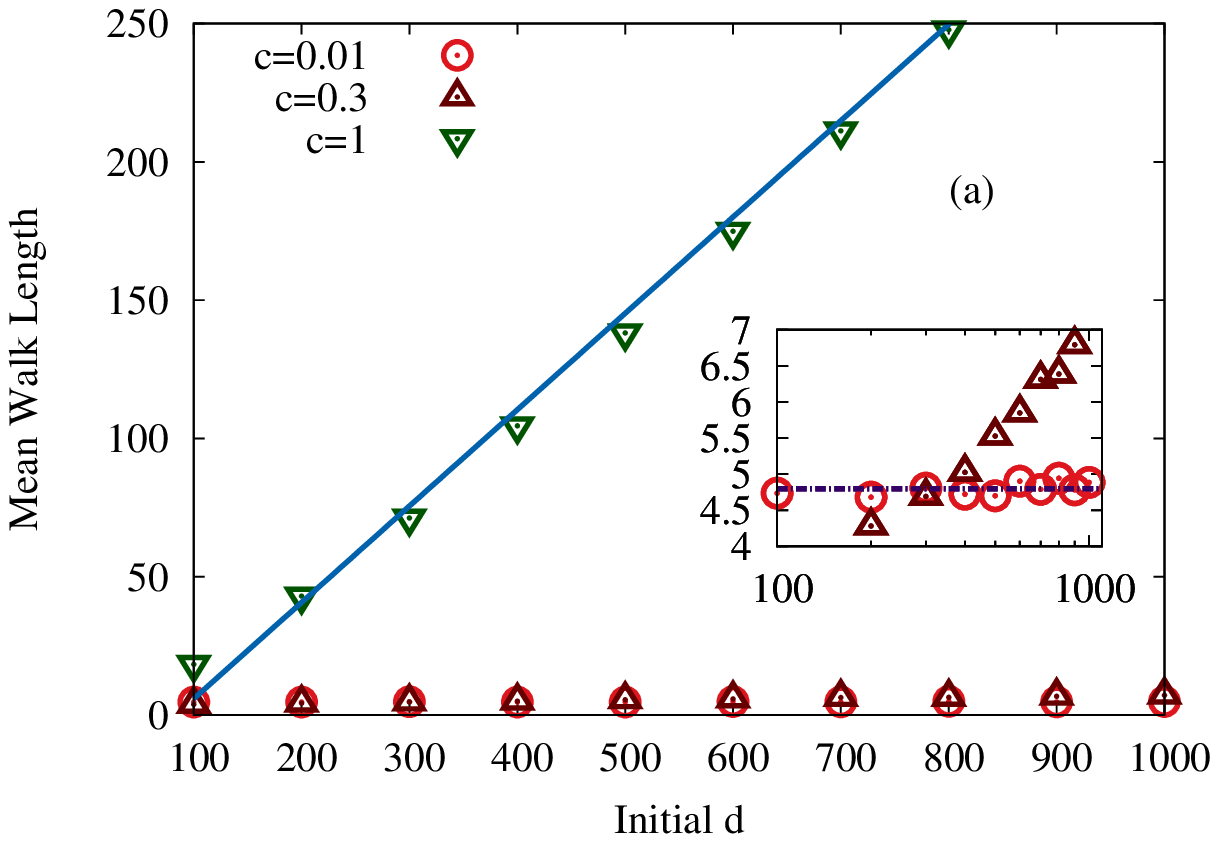}
	\end{minipage}
	\begin{minipage}{.49\textwidth}
		\includegraphics[width=\textwidth]{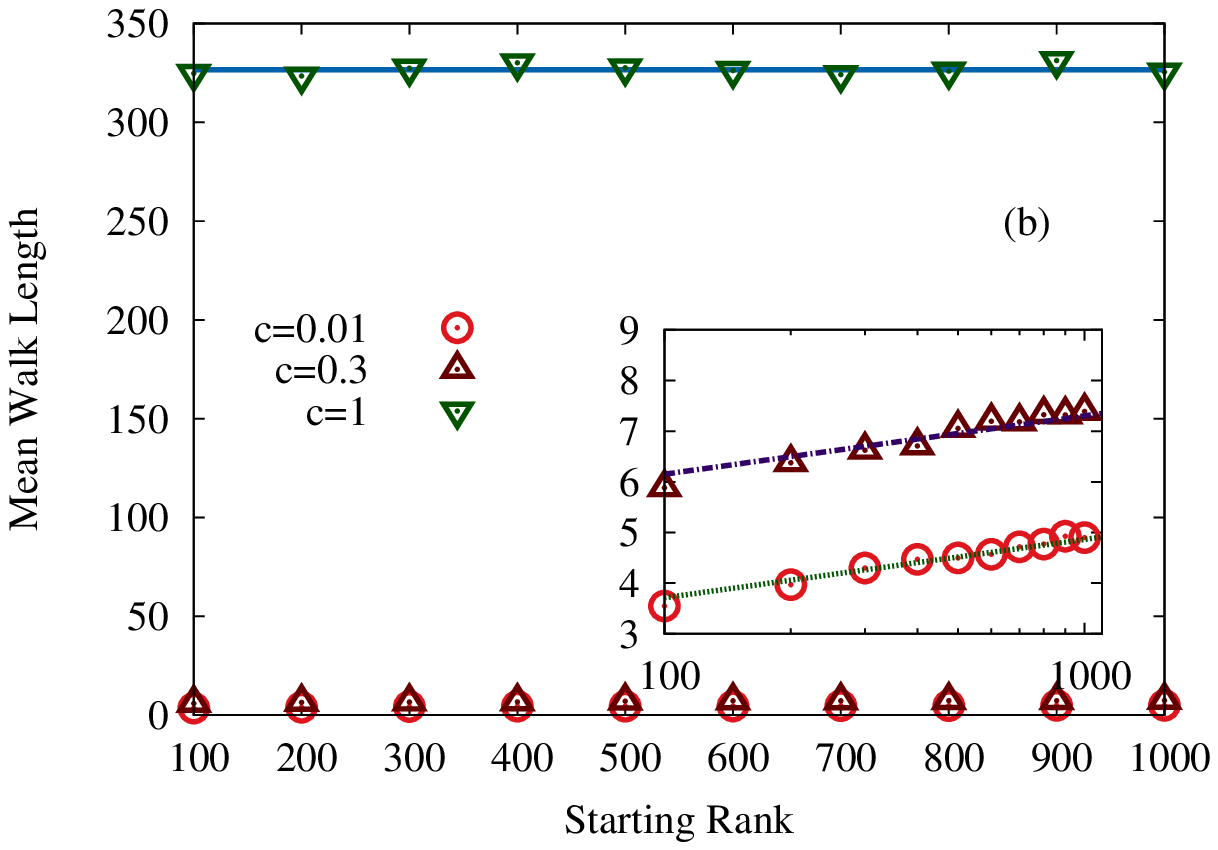}
	\end{minipage}
\caption{Mean length of adaptive walks in RMF landscapes with Gaussian randomness. (a) Mean walk length
for randomly chosen starting rank 
versus initial Hamming distance $d$ to the reference
sequence. Straight line illustrates the linear dependence of the
walk length on $d$ for large $c$. (b) Mean walk length for constant initial Hamming
distance $d=1000$ versus starting rank $r$. The horizontal line connecting
the data points for $c=1$ illustrates that walk length becomes independent of
starting rank for large $c$. Insets show the data for small
$c$ on logarithmic scales for $d$ and $r$, respectively. 
Horizontal line in the inset in panel (a) illustrates that the walk
length is independent of initial distance $d$ for $c=0.01$, but
acquires such a dependence with increasing $c$. Straight
lines in the inset of panel (b) illustrate the logarithmic dependence of the walk
length on initial rank for small $c$. The number of loci is $L=2000$.}
\label{fig:mfdr}
\end{figure}

Numerical results obtained from simulations of the full fitness-dependent AW
with transition probability in Equation \ref{eq:transprob} are shown in Figure \ref{fig:mfdr}.
These simulations were carried out for $L=2000$ loci and the random component of the fitness was drawn from a
normal distribution with unit variance. While in
the left panel the starting rank was kept constant, the initial Hamming distance to $\sigma^\ast$, $d$,  was varied. For
$c=0.01$ (inset), the behavior seems to be independent of $d$, while a $d$-dependence starts to emerge for $c=0.3$. 
For $c=1$ the relation between initial distance $d$ and the walk length is roughly linear with a slope smaller than unity.

For constant $d=\frac L 2 = 1000$ 
and various choices of the starting rank, one can observe the following behavior (Figure \ref{fig:mfdr} (b)). 
While for $c=1$ the walk length seems to be independent of the starting rank, the data for $c=0.01$ can be fitted by 
the relation
\begin{equation}
\label{eq:Gumbelwalk}
\ell = \frac 1 2 \log(r) + \text{const.}
\end{equation}
which was first obtained by \citeN{orr1} for the HoC landscape with Gumbel-distributed fitness values.    
For $c=0.3$ the dependence on the starting rank is similar but the
constant in Equation \ref{eq:Gumbelwalk} is significantly larger. 

\begin{figure}
	\begin{minipage}{.5\textwidth}
		\includegraphics[width=\textwidth]{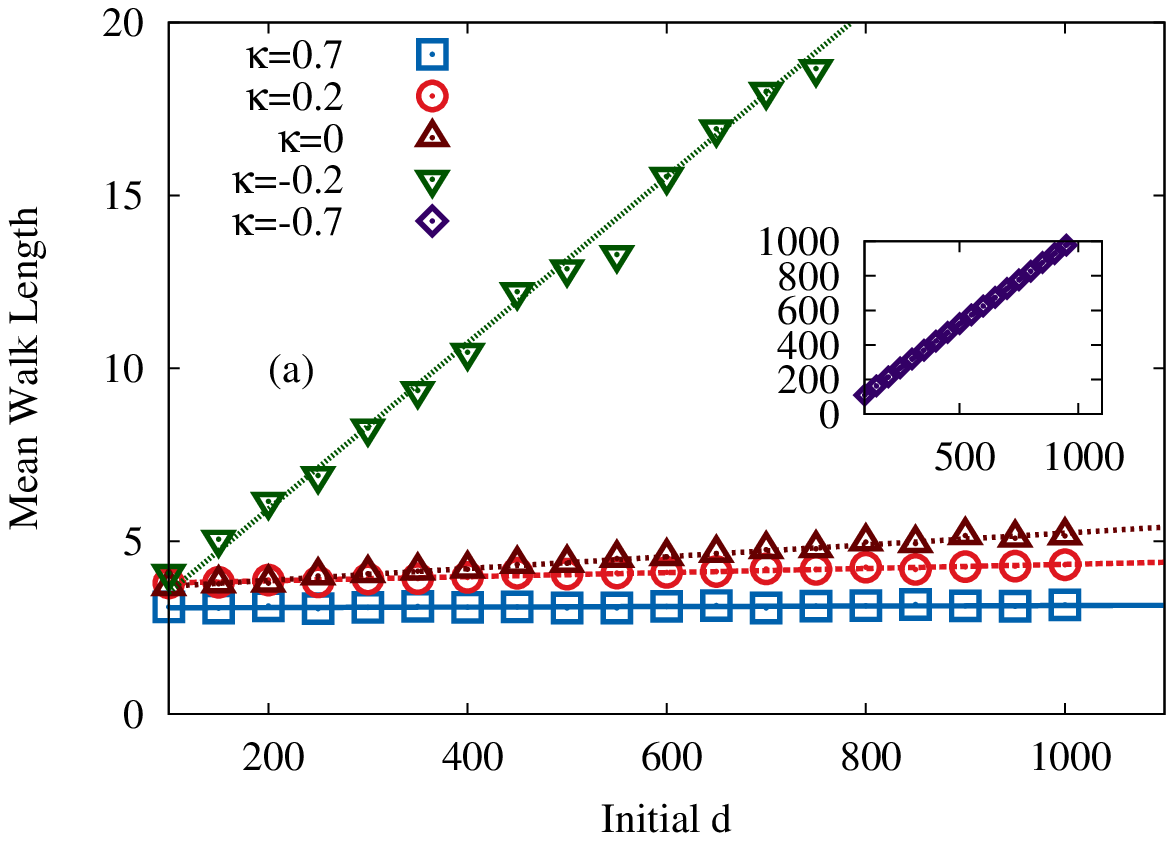}
	\end{minipage}
	\begin{minipage}{.5\textwidth}
		\includegraphics[width=\textwidth]{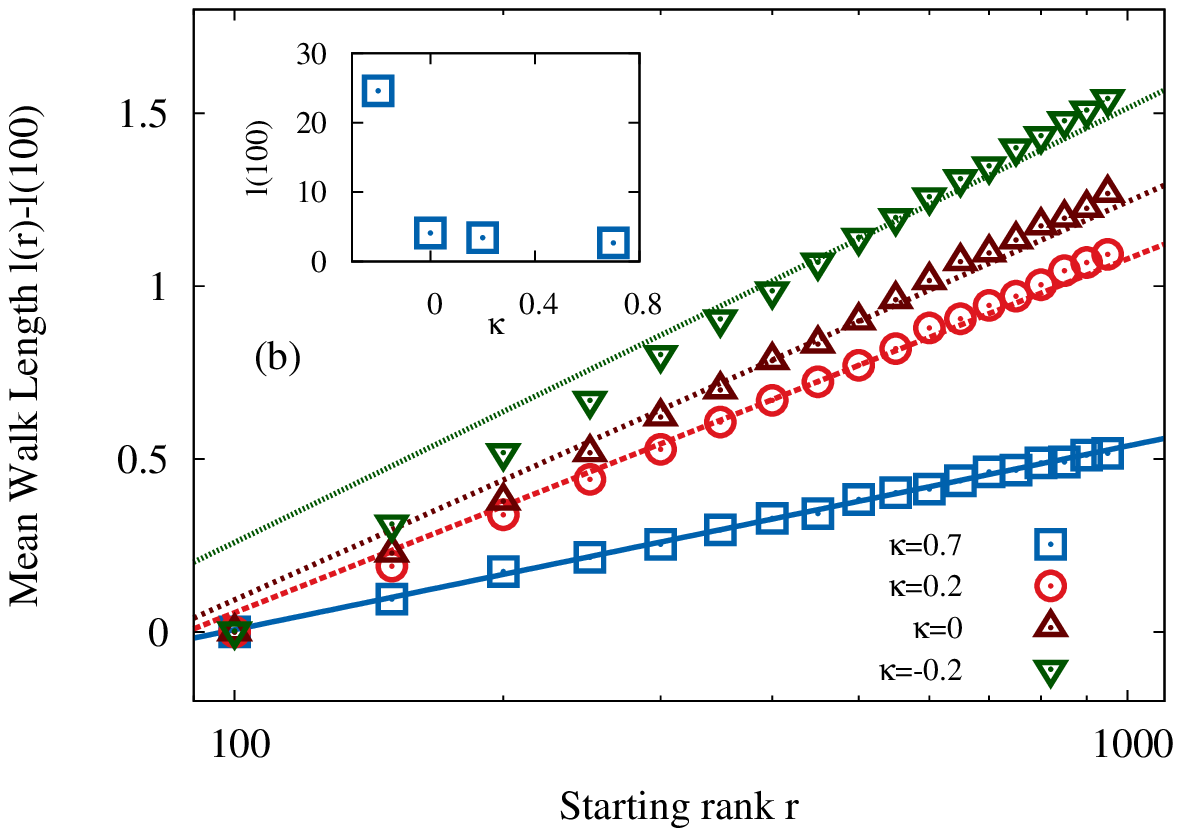}
	\end{minipage}
\caption{Mean length of adaptive walks in GPD distributed RMF
  landscapes for various choices of the EVT index $\kappa$. In both panels 
$c=0.5$. (a) Walks with
randomly chosen starting rank  and varying initial Hamming distance
$d$ to the reference sequence. Inset shows results for $\kappa = -0.7$, which
are off the scale of the main panel. For negative $\kappa$ (Weibull class) the walk length
displays a pronounced linear dependence on $d$. (b) Walks starting at constant Hamming 
distance $d=1000$ and varying starting rank $r$. In the main panel the
walk length for $r=100$ has been subtracted for clarity, and the
corresponding values of $\ell(100)$ are shown in the inset.
The lines in (a) correspond to fits assuming a linear $d$-dependence,
lines in (b) show the HoC result in Equation \ref{eq:walklength}. The latter provides a reasonable
fit to the data, if a $c$-dependent constant offset is allowed for, but the quality of the
approximation gets worse for negative $\kappa$ (Weibull class). The number of loci is $L=2000$.}
\label{fig:mfgpd}
\end{figure}

Equation \ref{eq:Gumbelwalk} is a special case of the general relation
\begin{align}
\ell=\frac{1-\kappa}{2-\kappa}\log(r)+\text{const.}\label{eq:walklength}
\end{align}
derived by \citeN{krugneid} and \citeN{Jain2011}, which holds for HoC landscapes with fitness distributions characterized
by an EVT index $\kappa \leq 1$; for $\kappa > 1$ the mean walk length is asymptotically independent of starting rank.
To see how this behavior is modified in the RMF model,  
we carried out simulations with the random fitness component drawn from the GPD, keeping $c=0.5$ fixed and varying
the EVT index $\kappa$, see Figure \ref{fig:mfgpd}. 
The $d$-dependence of the data seems to be well fitted by functions linear in $d$,
with a slope that increases rapidly with decreasing $\kappa$ when $\kappa < 0$; as was pointed out previously, the effect of the mean fitness gradient is particularly pronounced for 
distributions in the Weibull class. Somewhat surprisingly,  
the dependence on the starting rank at fixed $d$ can be approximately
described by the functional form in Equation \ref{eq:walklength}
obtained for the HoC model but with a constant depending on $c$ and
$\kappa$, see Figure \ref{fig:mfgpd} (b).

Summarizing, inspection of the numerical data suggests the following dependencies of the mean adaptive walk length:
a linear dependence on $d$ with a slope that increases with increasing $c$ and decreasing $\kappa$, and
a logarithmic dependence on the starting rank, similar to that known from the HoC model, with a constant
offset depending on $c$, $\kappa$ and $d$. These findings are captured  
in the following conjectured expression for the adaptive walk length on RMF landscapes: 
\begin{align}
\label{eq:walklength_gen}
\ell(r,c,d, \kappa)=\frac{1-\kappa}{2-\kappa}\log(r)+\alpha(c,\kappa) d + \beta(c,\kappa)
\end{align}
with so far unknown, nonlinear functions $\alpha, \beta$ with 
$\alpha(0,\kappa) = 0$ and $\beta(0,\kappa) > 0$ (see \citeN{krugneid} for a discussion of the $\kappa$-dependence of the 
constant term in Equation \ref{eq:walklength}).  

\subsection{Crossing Probability.}

\begin{figure}
    \begin{center}
    \includegraphics[width=0.6\textwidth]{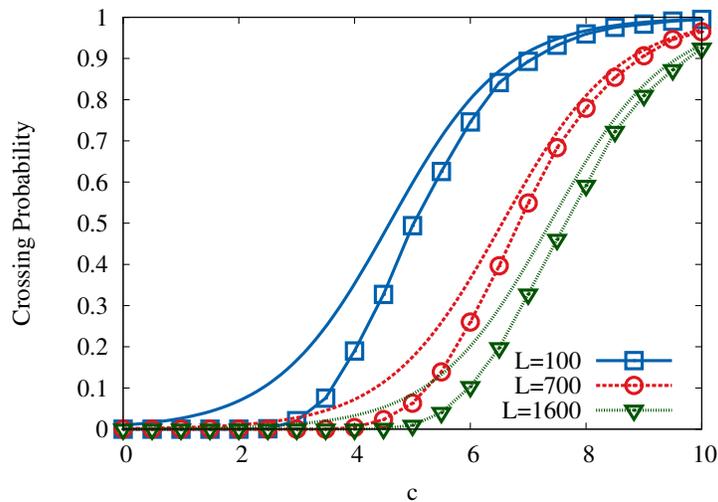}
    \caption{The probability for an adaptive walk starting from the
      antipodal sequence $\overline{\sigma^\ast}$ to reach the reference
      sequence $\sigma^\ast$ and terminate there. Results are shown
      for Gumbel-distributed random fitness components. Numerical
      results are displayed by symbols connected with lines, while the
    corresponding lines without symbols show the upper bound given in
    Equation \ref{pc0}.}
    \label{fig:pcross}
    \end{center}
\end{figure}

While the adaptive walk length is a measure of the length of \textit{typical} adaptive trajectories, it is also of interest to ask
how likely it is for the population to traverse the entire landscape. To quantify this
feature we introduce the \textit{crossing probability}, which is the probability that an AW
starting at the maximal distance $L$ from the reference sequence $\sigma^\ast$ reaches it and
terminates there. For such an event to happen, three conditions must
be fulfilled: The reference sequence $\sigma^\ast$ must be a
local maximum; there must exist at least one fitness-monotonic path
connecting the antipodal sequence $\overline{\sigma^\ast}$ to
$\sigma^\ast$; and finally, such a path must be chosen by the AW. 
The probability for the first condition was evaluated above and is
given by 
\begin{equation}
\label{pc0}
p_c^\mathrm{max}(0) = \frac{1}{1 + L e^{-c}}
\end{equation}
for Gumbel-distributed random fitness component, and this obviously
constitutes an upper bound on the crossing probability. The probability for
the existence of fitness-monotonic pathways in the RMF model has been investigated
previously for the case when the paths end at the global fitness
maximum \cite{jaspjoa}, and it has been shown that such paths exist with unit
probability for large $L$ and any $c > 0$ \cite{Hegarty2014}. It is
not clear whether this result applies in the present setting, however,
because the probability that the global maximum coincides with the
reference sequence vanishes for large $L$ (see Equation \ref{global_dist}). 
Figure \ref{fig:pcross} shows numerical data for the crossing
probability in comparison with the upper bound in Equation \ref{pc0}. Both
quantities follow a sigmoidal behavior with a fairly sharp transition
from zero to unity around a characteristic value of $c$, which
increases roughly logarithmically with $L$, as would be expected on
the basis of Equation \ref{pc0}.   

\section{The Number of Exceedances}

In this section we consider a feature of the fitness landscape that provides a distribution-free 
statistical test of the assumption of the MLM that fitness values of different genotypes
are i.i.d. random variables \cite{miller}. To define the quantity of interest, 
suppose that an adaptive step is taken from a starting genotype $\sigma$ with rank $i$ in its 
neighborhood $\nu(\sigma)$ to a genotype $\sigma^\prime$ with rank $j < i$ in $\nu(\sigma)$. 
The \textit{number of exceedances (NoE)} is then equal to 
the number of neighboring genotypes in $\nu(\sigma')$ that are fitter than $\sigma'$, that is,
the rank of $\sigma'$ in its own neighborhood minus one. Since the only sequences present in both
neighborhoods $\nu(\sigma)$ and $\nu(\sigma^\prime)$ are $\sigma$ and $\sigma^\prime$,
the NoE can principally vary between $0$ and $L-1$. Under the HoC assumption that fitness values
are i.i.d. random variables, a classic result due to \citeN{gumbel3} states that the distribution
of the NoE is independent of the fitness distribution, and for large $L$ the mean NoE is equal to the 
rank of $\sigma'$ in the initial neighborhood $\nu(\sigma)$ \cite{rokyta}. In other words,
if the first adaptive step goes to a genotype of rank $j$, the expected number of 
beneficial mutations available for the second step is $j$. In contrast, in a purely additive landscape
such as the RMF model for very large $\theta$, the NoE of a genotype is equal to its
distance $d$ from the global fitness maximum and independent of its rank in the initial neighborhood.  

In their evolution experiments with the microvirid bacteriophage ID11, \citeN{miller} identified
9 beneficial second step mutations on the background of a mutation, named g2534t, that had been
found to have the largest effect among 16 beneficial first step mutants. In order to apply the result of 
\citeN{gumbel3}, the rank of this mutation among \textit{all} possible first step mutations (not only
the 16 observed in the experiment) has to be estimated.   
Assuming conservatively that the rank of g2534t among all beneficial first step mutations is at most 3,
at most 3 beneficial second step mutations would then have
been expected if fitness values were identically and independently distributed. 
Thus, the observation of 9 beneficial second step mutations allowed \citeN{miller} to reject the HoC hypothesis
with high confidence ($P < 0.02$). 

Here we ask whether the RMF model is capable of yielding predictions for the NoE 
that are compatible with the experimental findings of \citeN{miller}. Note that, unlike HoC landscapes, RMF landscapes are not isotropic and the NoE will depend on the position of genotypes $\sigma$ and $\sigma'$ on the landscape, i.e.\ their distance to the reference state, and on whether 
the adaptive step was taken in the uphill or downhill direction. In contrast to the universal
result of \citeN{gumbel3}, there will also be a dependence on the probability
distribution of fitness values. 

\begin{figure*}
    \begin{minipage}{.49\textwidth}
    \includegraphics[width=\textwidth]{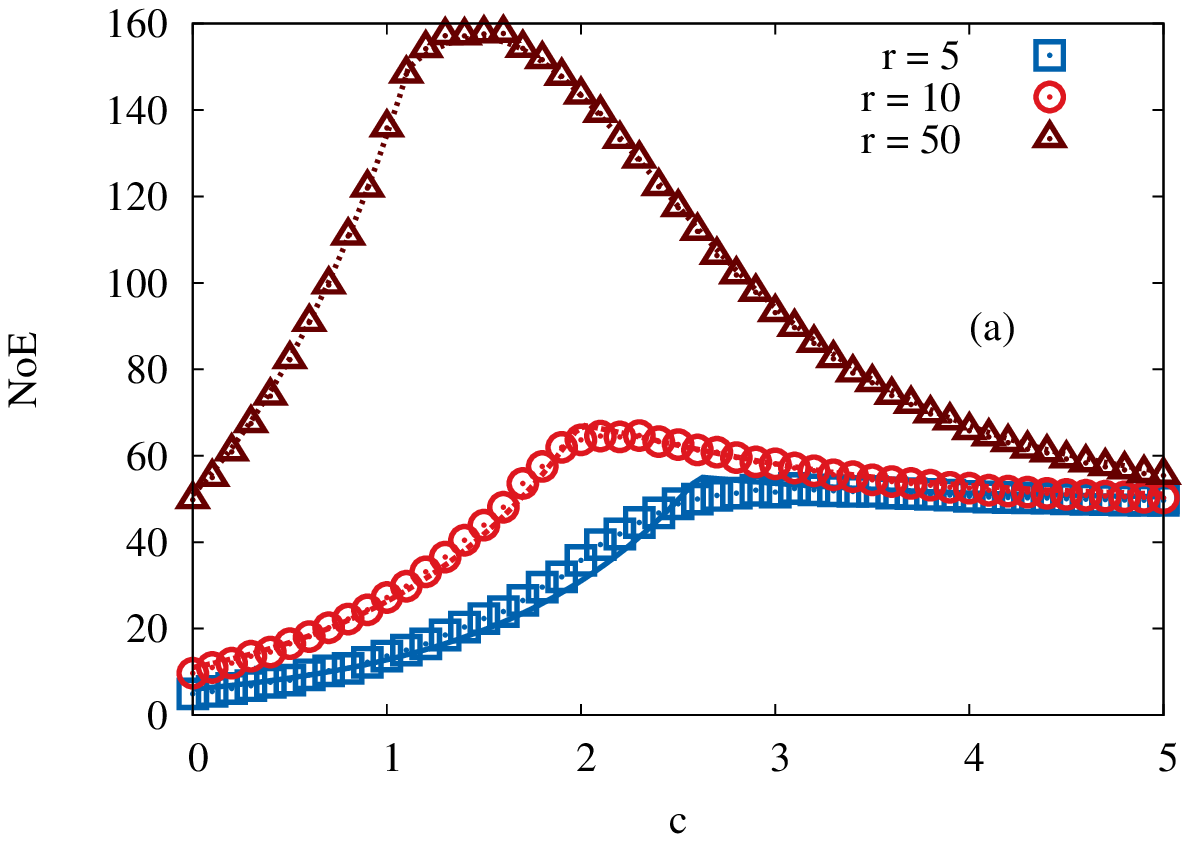}
    \end{minipage}
    \hspace{.5cm}
    \begin{minipage}{.49\textwidth}
    \includegraphics[width=\textwidth]{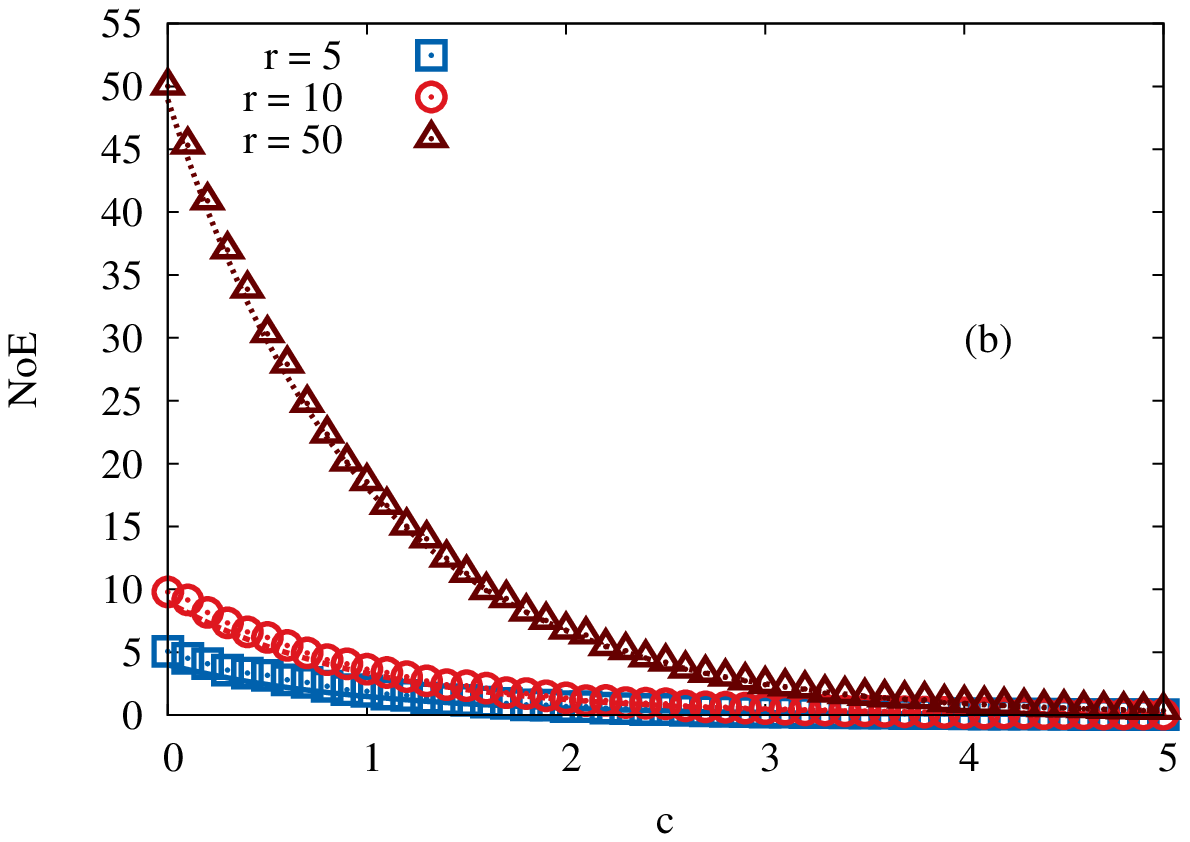}
    \end{minipage}
    \caption{Expected number of times that a sequence which had rank $r$ in the old neighborhood 
is exceeded in fitness in the new neighborhood, when the adaptive step occurred in the (a) uphill
or (b) downhill direction. Symbols show simulation results for the RMF model with exponentially
distributed random fitness component, and lines show the approximate analytic expression 
derived in APPENDIX C. In all cases the initial genotype was located at distance $d=50$ from
the reference sequence in a landscape with $L = 1000$ loci. The limiting values of the NoE
are ${\cal{N}}^\mathrm{up} = {\cal{N}}^\mathrm{down} = r$ for the HoC-model ($c=0$) and ${\cal{N}}^\mathrm{up} = 
d$, ${\cal{N}}^\mathrm{down} = 0$ for a smooth landscape ($c \to \infty$).}
    \label{fig:stepupdown}
\end{figure*}

In APPENDIX C we present an approximate analytic calculation of the expected 
NoE for the RMF model, assuming
an exponential distribution for the random fitness component. While the complete expressions 
for the expected NoE displayed in Equations \ref{kgt2} - \ref{Ndown} are fairly complex, for small 
$c$ they reduce to the simple form
\begin{equation}
\label{NoE_simple} 
{\cal{N}}^\mathrm{up} \approx 2 + (r-1)e^c, \;\;\;\; {\cal{N}}^\mathrm{down} \approx 2 + (r-1)e^{-c}.
\end{equation}
Here ${\cal{N}}^\mathrm{up}$ (${\cal{N}}^\mathrm{down}$) is the expected number of exceedances
after an adaptive step in the uphill (downhill) direction, and $r$ is the rank of the mutated genotype
$\sigma'$ in the initial neighborhood. For the HoC landscape ($c = 0$)
Equations \ref{NoE_simple} yield ${\cal{N}}^\mathrm{up} = {\cal{N}}^\mathrm{down} = r + 1$, which differs slightly from the 
exact result ${\cal{N}} = r$ as a consequence of the approximations involved in the derivation. 
Figure \ref{fig:stepupdown} compares the full expressions derived in APPENDIX C to numerical 
simulations, showing good agreement. Interestingly, for the case of an uphill step, the expected
number of exceedances is maximal for landscapes of intermediate ruggedness.

\begin{figure*}
     \begin{minipage}{0.49\textwidth}
    \includegraphics[width=\textwidth]{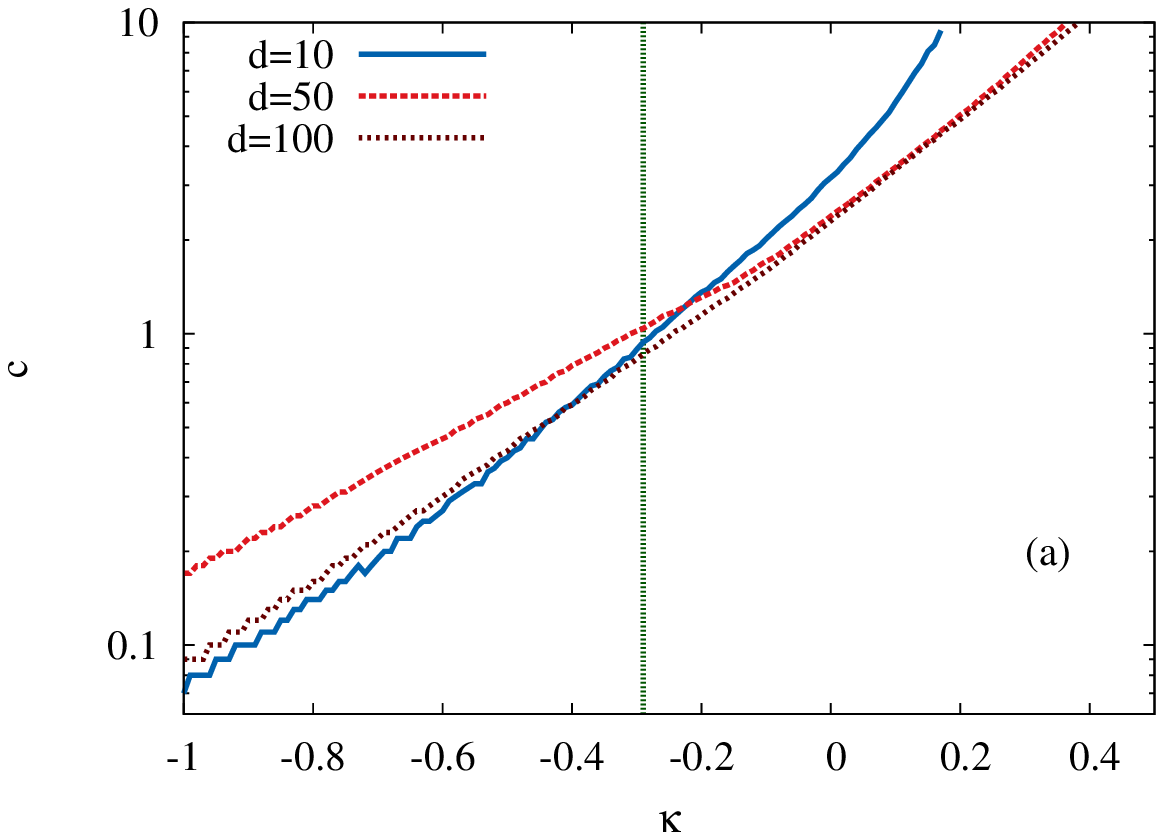}
     \end{minipage}
     \hspace{.5cm}
     \begin{minipage}{.49\textwidth}
     \includegraphics[width=\textwidth]{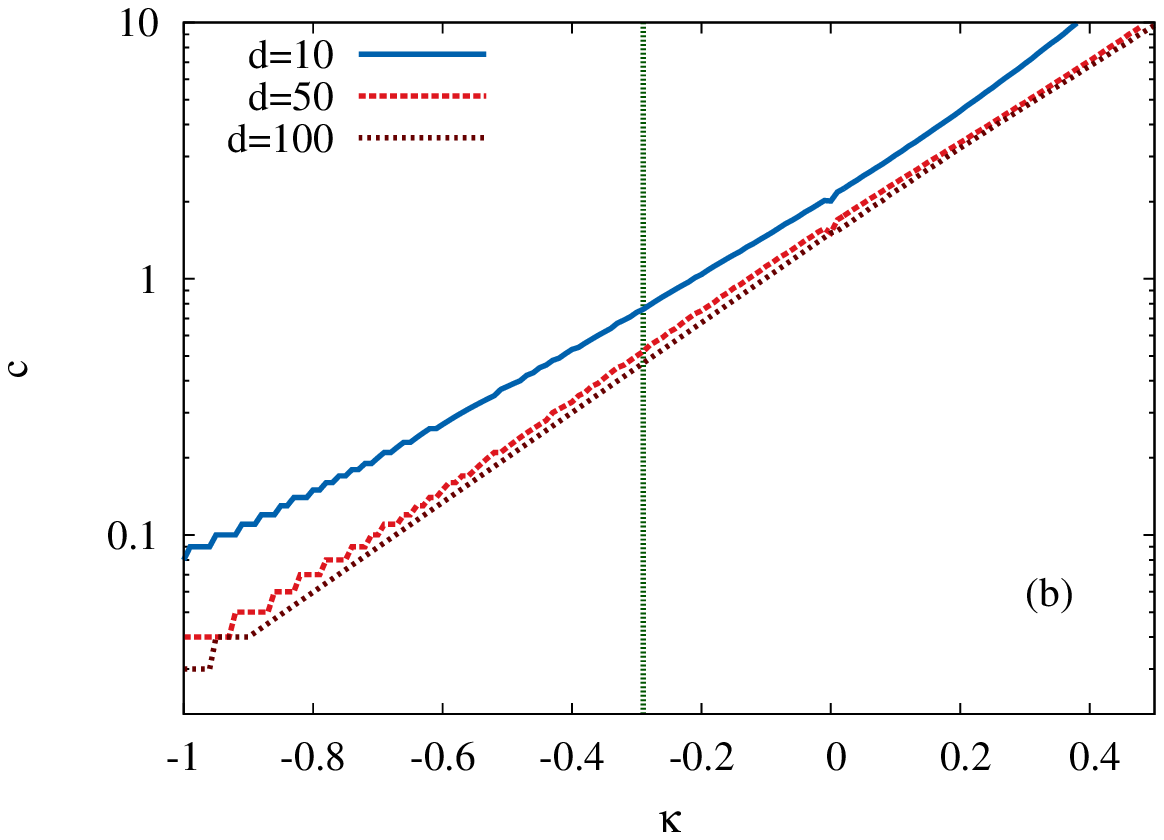}
     \end{minipage}
\caption{The figure shows the minimal value of the RMF parameter $c$
  required to generate on average 9 exceedances after an adaptive
  step. Panel
  (a) shows results for initial rank 1 and panel (b) for initial rank 3.
  The random
  fitness component is assumed to be distributed according to the GPD
  distribution with EVT index $\kappa$, and different curves
  correspond to different values of the initial distance $d$ to the
  reference sequence. The experimental estimate $\kappa \approx -0.29$
  of the EVT index is indicated by a vertical line. The total number of loci is $L=1000$.
Figure \ref{fig:csigmexc} shows the same results in a plot where the additive selection
coefficient has been scaled by the standard deviation of the distribution of the random
component.}
\label{fig:expsupp}
\end{figure*}

Equation \ref{NoE_simple} shows that a considerable enhancement of
the NoE is possible for moderate $c$, provided the adaptive step is in
the uphill direction and the random fitness components are
exponentially distributed. 
However, as we do not know whether the beneficial mutations that were
observed in the experiments of \citeN{miller} correspond to uphill or
downhill steps in a presumed RMF landscape, we need to average the predictions for the two cases,
weighted with the probabilities for each of the two types of transitions to have happened. 
Furthermore, for an unbiased comparison it is appropriate to consider more general distributions for
the random component than the exponential distribution discussed above. Here we choose the GPD distribution, which allows us to cover distributions corresponding to all
extreme value classes by varying a single parameter. Extending
the approximate derivation in APPENDIX C to this general case is
complicated and not very enlightening. 
Therefore, in the following we use numerical simulations to estimate
the NoE. 

To find parameter combinations that match the experimentally observed
NoE, we constructed RMF landscapes fixing the GPD index in the interval
$\kappa \in [-1,0.5]$, and sampling $\kappa$ with a resolution of
$\Delta \kappa = 0.01$. For each value of $\kappa$, the NoE was
calculated for a range of $c$ values. Following \citeN{miller} we
assume that the first step mutation has rank $r=1$ in the
initial neighborhood, and determine for each choice of $\kappa$ the
smallest value of $c$ for which $\mathcal N(1) \geq 9$
(Figure \ref{fig:expsupp} (a)). For comparison, 
results assuming initial rank $r=3$ are shown in Figure
\ref{fig:expsupp} (b). 
Note that, similar to the case of the exponential distribution
displayed in Figure \ref{fig:stepupdown}, for some $\kappa$ there exists a second, 
larger $c$ yielding the same value for $\mathcal N$. 
The results displayed in Figure \ref{fig:expsupp} show that the
strength of the fitness gradient $c$ required to reproduce the
experimentally observed NoE depends sensitively on the EVT index, and
becomes very small for negative $\kappa$ deep inside the Weibull
domain. 

This is in accordance with the generally stronger effect of the additive
fitness contribution for negative $\kappa$ discussed above in the
context of the statistics of fitness maxima, and reflects the fact
that for strongly negative $\kappa$ the random variables drawn from the 
GPD probability density tend to crowd near the upper boundary of its
support. For the EVT index $\kappa = -0.29$ estimated by
\citeN{miller} from a maximum likelihood analysis of the fitness
values of 16 first step beneficial mutations, the $c$-value required
to produce at least 9 exceedances varies between 0.86 and
1.04 for initial rank 1 and between 0.44 and 0.76 for initial rank 3,  
depending on the assumed distance $d$ to the reference sequence. 
Normalizing $c$ by the standard deviation of the random fitness component which
equals 0.617 for this value of $\kappa$ (compare to Equation \ref{GPD_StD}), this translates
into the intervals $1.39 \leq \theta \leq 1.69$ and $0.71 \leq \theta \leq 1.23$
for the parameter $\theta$, respectively, which in comparison to estimates for 
other empirical data sets is indicative of a relatively smooth landscape.   
To further narrow down the range of RMF model parameters that
can provide a consistent description of the fitness landscape of the
ID11 microvirid system would require including other experimental
observables into the analysis, which is beyond the scope of this article. 

\section{Discussion}

Motivated by the increasing availability of empirical information about the structure of 
adaptive landscapes, we have presented a detailed analysis of a one-parameter family of 
tunably rugged fitness landscapes. The model is a variant of the Rough Mount Fuji
landscape, in which the locus-specific additive effects considered in the original version 
\cite{aita,Aita2000} are replaced by a single parameter, and ruggedness is governed by 
the ratio $\theta$ of the additive fitness effect to the standard deviation of the random fitness component.

\subsection{Landscape topography.} The mathematical
simplicity of the model allowed us to derive several explicit results for  
the number and positions of local and global maxima; such results are much harder to come by
for other generic fitness landscape models, notably Kauffman's NK-model. 
In particular, we have arrived at a complete classification of the expected
number of fitness maxima in the limit of large genotype dimensionality $L$, which highlights
the importance of the tail behavior of the distribution of the random fitness component
in determining the ruggedness of the landscape. For distributions with tails heavier than
exponential, the additive fitness component becomes asymptotically irrelevant, in the
sense that the number of maxima is equal to the value expected for an uncorrelated HoC landscape.
In contrast, for distributions with bounded support the number of maxima is reduced compared
to the HoC expectation by a factor that varies exponentially in $L$. The interplay between
additive effects and tail behavior has been a recurrent theme of this article which manifests
itself in various quantities of interest. Our results show that both the parameter
$\theta$ characterizing the additive effects and the EVT index $\kappa$ must be specified for a comprehensive description of the landscape topography in the RMF model. 

An exception to this rule is provided by the fitness correlation function which, 
similar to the NK-model, is independent of the type of randomness. In contrast to the NK-model,
however, the correlations become negative at large distances, which reflects the inherent 
anisotropy of the RMF landscape and the long-range effect of the fitness gradient.

Another important measure of fitness landscape ruggedness not discussed so far in this article is the
existence and abundance of selectively accessible mutational pathways, defined as paths composed 
of single mutational steps along which fitness increases monotonically \cite{Weinreich2006,jaspjoa}. 
Using an approach similar to that of the present work, an explicit expression for the expected number
of accessible pathways in the RMF model with Gumbel-distributed randomness can be derived 
\cite{jaspergregor,jaspjoa}. Subsequently \citeN{Hegarty2014} presented a rigorous proof that
accessible pathways exist with unit probability for large $L$ in the RMF model for any $c > 0$, 
independent of the distribution of the random fitness component. This is in stark contrast to the 
behavior in the HoC model ($c=0$), where the probability for existence of accessible paths tends
to zero for large $L$. Analyses in which the genotypes are placed 
on a regular tree show that this strong dichotomy between the HoC and RMF models is specific to 
the hypercube topology of sequence space \cite{Nowak2013,Roberts2013}.  

\subsection{Dynamics of adaptation.}
Apart from being amenable to rigorous analysis, the RMF model is useful for exploring
various aspects of evolutionary dynamics in rugged fitness landscapes through simulations. 
Recent applications in this context include studies of evolutionary predictability 
\cite{Lobkovsky2011,Szendro2013a}, epistatic interactions between mutations 
occurring along an adaptive walk \cite{Greene2014} and the effect of recombination
on rugged fitness landscapes \cite{Nowak2014}. 
Here we have focused on adaptation in the SSWM regime
and considered both single adaptive steps and adaptive walks to local fitness maxima. Interestingly,
while the statistics of single adaptive steps largely conforms to the classic results obtained for
the MLM \cite{orr1,joyceorr}, adaptive walks in the RMF are much
longer than in the MLM already for small values of $c$. Specifically,
the heuristic expression in Equation \ref{eq:walklength_gen} that summarizes the
simulation results suggest a linear dependence of the walk length on
the initial distance to the reference sequence. 

The qualitatively
different effects that the fitness correlations in the RMF have on
single vs. multiple adaptive steps highlight the fact that a step in
an adaptive walk involves two distinct random processes
\cite{rokyta,krugneid}. The first process is the selection of a fitter
neighbor according to the transition probability in Equation \ref{eq:transprob}, and the
second process is the change of the mutational neighborhood after the
fixation of the mutated genotype. In the MLM the effect of the second
process is relatively weak, and as a consequence adaptive walks are
well described by an approximation which ignores the neighborhood
change \cite{orr1,krugneid}. 

To understand the role of the neighborhood 
change in the RMF model we refer to the analysis in APPENDIX C, where
it is shown (for exponentially distributed randomness) that the effect
of the fitness gradient can be approximately subsumed into an overall
shift of \textit{all} the fitness values constituting a neighborhood by the same
amount. This implies that the transition probability in Equation \ref{eq:transprob}, which
depends only on fitness differences within the neighborhood, is
approximately independent of $c$. However, since the shift is a
function of the distance $d$ to the reference sequence which changes in the
adaptive step, the rank of the mutant genotype in the new neighborhood
is strongly dependent on whether the step occurred in the uphill ($d
\to d-1$) or downhill ($d \to d+1$) direction. Further investigations are 
required to elucidate how this effect gives rise to the observed dependence 
of the walk length on the landscape parameters. A promising approach is to consider 
walks starting from the antipode of the reference sequence, which can be assumed to
move almost exclusively in the uphill direction provided the walk length remains short 
compared to $L$ \cite{Park2014}.

For large populations and/or high mutation rates the SSWM approximation underlying the simple
adaptive walk picture breaks down, and additional processes such as the competition between 
beneficial clones and the crossing of fitness valleys have to be taken into account. The resulting complex
population dynamics is governed by the tension between a tendency towards greater determinism
induced by clonal competition \cite{Jain2007,Jain2011a} and the increasing role of non-monotonic pathways
that become accessible through valley crossing \cite{Szendro2013a,deVisser2014}. In contrast to the SSWM regime,
the trapping of large populations at local fitness maxima is only a transient occurrence, and the rate
of escape from such peaks plays an important role in the comparison between recombining and non-recombining
populations on rugged fitness landscapes \cite{Nowak2014}. 

\subsection{Application to experiments.} The usefulness of the RMF model
for the quantitative description of empirical fitness landscapes has
been documented in several recent studies. \citeN{jaspjoa} applied the
model to an 8-locus fitness landscape for the fungus \textit{Aspergillus niger}, and
extracted an estimate of $c$ from a subgraph analysis of
pathway accessibility. In a study of amplitude spectra of fitness
landscapes \citeN{Neidhart2013} showed that the correlation
function of a 6-locus fitness landscape obtained by \citeN{Hall2010}
for the yeast \textit{Saccharomyces cerevisiae} is
well described by the RMF model. Lastly, in a meta-analysis of 10
empirical fitness landscapes \citeN{Szendro2012} used the RMF model to
interpolate the behavior of various ruggedness measures between the
limits of a completely random (HoC) and an additive landscape, and
found good agreement with the trends in the empirical data.    

In the present article we have complemented these analyses by
considering the effect that the fitness gradient in the RMF model has
on the number of secondary beneficial mutations that are available
after an adaptive step. We have identified model parameters for which
the RMF prediction matches the large number of fitness exceedances
observed in the experiment of \citeN{miller}. A small
fitness gradient suffices to explain the experiments when the
distribution of the random fitness component is assumed to belong to
the Weibull class of EVT, as is suggested by the analysis of the
distribution of first-step mutational effects. We believe that more
work along these lines, focusing on the changes in the statistics of mutational
neighborhoods along an adaptive trajectory, will provide important
insights into the role of epistatic interactions during adaptation and
the viability of schematic models like the one considered here.  

\section{Acknowledgements}
We thank Su-Chan Park for discussions and two anonymous reviewers for 
hepful comments. JK acknowledges the kind hospitality of the Simons Institute for the Theory of Computing, Berkeley,
and the Galileo Galilei Institute for Theoretical Physics, Florence, during the completion of the paper. 
This work has been supported by DFG within SFB 680, SFB TR12, SPP 1590 and 
the Bonn-Cologne Graduate School of Physics and Astronomy.


\appendix
\renewcommand{\theequation}{A\arabic{equation}}
\renewcommand{\thefigure}{A\arabic{figure}}
\setcounter{equation}{0}
\setcounter{figure}{0}
\section{Appendix A: Properties of fitness maxima}
\label{app:A}

\subsection{Density of local maxima.}
The probability $p^\text{max}_c(d)$ that a genotype at distance $d$
from the reference sequence is a local fitness maximum is given by the integral
\begin{align}
\label{eq:pmax(d)}
   p^\text{max}_c(d) &=\int dx \; p(x) \left(P(x-c)\right)^d \left(P(x+c)\right)^{L-d}.
\end{align}
This is simply the probability that the genotype's fitness $x$
exceeds that of its uphill and downhill neighbors, averaged with
respect to the probability density $p(x)$. Unless specified otherwise,
here and in the following
the domain of integration is equal to the support of the probability distribution.
For the Gumbel distribution $P_G(x)=e^{-e^{-x}}$ the integral 
\eqref{eq:pmax(d)} can be evaluated exactly using the shift property
\eqref{eq:Pshift}, yielding the result in Equation \ref{eq:pmaxgum} of the main text. 

Similarly, the probabilities to find the neighboring genotype of largest fitness
in the uphill or downhill direct, respectively, are given by  
\begin{align} \label{eq:pupdown(d)}
    p_c^\text{up}(d) &=d\int dx \; p(x) P(x)^{d-1}P(x+c)P(x+2c)^{L-d}, \\
    p_c^\text{down}(d) &=(L-d)\int dx \; p(x) P(x)^{L-d-1}P(x-c)P(x-2c)^{d},
\end{align}
which can be evaluated for the Gumbel distribution to yield the expressions 
in Equation \ref{eq:pupdown_Gumbel}.

\subsection{Total number of maxima.} Inserting Equation \ref{eq:pmax(d)}
into the sum in Equation \ref{eq:nomsum} and exchanging the order of integration and
summation one arrives at the compact expression 
\begin{equation}
\label{M:general}
{\cal{M}} = \int dx \; p(x) [P(x-c) + P(x+c)]^L
\end{equation}
which will be evaluated in the following for various special cases.
Equation \ref{M:general}  makes it evident that ${\cal{M}}$ is an even
function of $c$: Changing $c$ to $-c$ produces a fitness landscape
with the antipodal reference sequence $\overline{\sigma^\ast}$ 
that is statistically equivalent to the original landscape. 

\subsubsection{Exponential distribution.} For the exponential
distribution defined by 
\begin{align}
\label{eq:exponential}
     P(x) &=\begin{cases}
                            1-e^{-x} & x \geq 0\\
                            0 & x < 0
                         \end{cases}
\end{align} 
the expression in Equation \ref{M:general} becomes
\begin{equation}
\label{M:exp1}
{\cal{M}} = \int_0^c dx \; e^{-x} [1 - e^{-c} e^{-x}]^L + 2^L
\int_c^\infty dx \; e^{-x} [1 - \cosh(c) e^{-x}]^L. 
\end{equation}
Substituting $z = e^{-x}$ this yields
\begin{eqnarray}
\label{M:exp2}
{\cal{M}} = \int_{e^{-c}}^1 dz \; [1-e^{-c} z]^L + 2^L \int_0^{e^{-c}}
dz [1- \cosh(c) z]^L = \nonumber \\
\frac{e^c}{L+1} \left[ (1 - e^{-2c})^{L+1} - (1 - e^{-c})^{L+1}
\right] + 
\frac{2^L}{\cosh(c) (L+1)} \left[ 1 - (1-e^{-c} \cosh(c))^L \right].
\end{eqnarray} 
It is straightforward to check that ${\cal{M}} \to 1$ for $c \to \infty$.
Moreover, the asymptotics for large $L$ at fixed $c$ is identical to
that derived in Equation \ref{eq:largeL} for the Gumbel distribution,
${\cal{M}} \sim \frac{2^L}{L \cosh(c)}$. 

\subsubsection{Gumbel distribution.} Inserting the exact expression
in Equation \ref{eq:pmaxgum} into Equation \ref{eq:nomsum} one obtains 
\begin{equation}
\label{eq:MGumbel}
\mathcal M =\sum_{d=0}^{L}\binom{L}{d} \frac{1}{1+de^c + (L-d)e^{-c}}.
\end{equation} 
We first show that the sum in Equation \ref{eq:MGumbel} can be expressed in terms
of a hypergeometric function defined by \cite{concrete_math} 
\begin{align*}
   \,_2F_1(a,b;c;z) &= \sum_{n\geq 0} \frac{(a)_n(b)_n}{(c)_n} \,
   \frac {z^n} {n!} =\sum_{n\geq 0}t_n
\end{align*}
where the Pochhammer symbol is defined by
\begin{align*}
    (x)_n = \left\{ \begin{array}{ll}  1                     & \mbox{if } n = 0 \\  x(x+1) \cdots (x+n-1) & \mbox{if } n
    > 0. \end{array}\right. 
\end{align*}
The defining feature of the hypergeometric function is that the terms $t_n$ satisfy $t_0=1$ and 
\begin{equation}
\label{eq:tratios}
\frac{t_{k+1}}{t_k} = \frac{(k+a)(k+b)}{k+c}\frac{z}{k+1}.
\end{equation}
To bring Equation \ref{eq:MGumbel} into this form we write
\begin{equation} 
\mathcal M =\frac{1}{1+Le^{-c}} \sum_{d\geq 0} \binom{L}{d}
p^\text{max}_c(d)(1+Le^{-c}),
\end{equation}
ensuring that $t_0 = 1$, and compute the fractions $t_{d+1}/t_d$ according to 
\begin{equation}
\frac{t_{d+1}}{t_d}
=\left(\frac{L-d}{d+1}\right)\frac{1+Le^{-c} + 2d \sinh(c)}{1+Le^{-c}+2(d+1)\sinh(c)} =
\left(\frac{-1}{d+1}\right)\frac{(d-L)\left(d+\frac{1+Le^{-c}}{2\sinh(c)}\right)}{d+1+\frac{1+Le^{-c}}{2 \sinh(c)}}.
\end{equation}
By comparison with Equation \ref{eq:tratios} the arguments $a, b, c$ and $z$
can be identified and it follows that
\begin{equation}
\label{eq:nomsum_Gumbel}
\mathcal M = (1 + L e^{-c})^{-1} \; {_2F_1}(-L, \zeta; \zeta+1; -1)
  \;\;\; \text{with} \;\;\; \zeta = \frac{1+Le^{-c}}{2 \sinh c}. 
\end{equation}
The asymptotic behavior for large $L$ is most conveniently extracted
from the general expression in Equation \ref{M:general}, which in this case can be
brought into the form
\begin{equation}
\label{M:Gumbel}
{\cal{M}} = \int_0^1 dP \; [P^{e^c} + P^{e^{-c}}]^{L}.
\end{equation}
Recognizing that the dominant contribution to the integral comes from
the region near $P = 1$ and expanding the integrand around this point
one readily finds that ${\cal{M}} \approx \frac{2^L}{L \cosh(c)}$ for
large $L$, in agreement with Equation \ref{eq:largeL}. Along similar lines it can 
be shown that the same asymptotics obtains for any distribution with a strictly
exponential tail. 

\subsubsection{Fr\'echet class.} Here we show that for distributions
in the Fr\'echet class the expected number of
maxima is asymptotically equal to that in the HoC model, ${\cal{M}}
\approx \frac{2^L}{L}$, for large $L$ and any $c > 0$. To simplify the
notation we use a standard Pareto distribution defined by 
\label{eq:Pareto}
\begin{align}
     P(x) &=\begin{cases}
                            1-x^{-\alpha} & x \geq 1\\
                            0 & x < 1
                         \end{cases}
\end{align} 
where the exponent $\alpha$ is related to the EVT index through
$\alpha = \frac{1}{\kappa}$. Similar to the case of the exponential
distribution, the domain of integration in Equation \ref{M:general} has to be subdivided 
into the interval $[1,1+c)$, where $P(x-c) = 0$, and the remainder
$[1+c,\infty)$. The dominant contribution comes from the second part,
which is given by
\begin{eqnarray}
\label{M:Pareto1}
\mathcal M \approx 2^L \int_{1+c}^\infty dx \; \alpha x^{-(\alpha +
  1)} \left[ 1 - \frac{1}{2}[(x+c)^{-\alpha} + (x - c)^{-\alpha}]
\right]^L \nonumber \\
\approx 2^L \int_{1+c}^\infty dx \; \alpha x^{-(\alpha +
  1)} \exp \left[- \frac{1}{2} L x^{-\alpha} [(1 + c/x)^{-\alpha} + (1 -
  c/x)^{-\alpha}] \right]
\end{eqnarray} 
for large $L$. Substituting $y = L x^{-\alpha}$ this yields
\begin{equation}
\label{M:Pareto2}
\mathcal M \approx \frac{2^L}{L} \int_0^{\frac{L}{(1+c)^\alpha}} dy \; 
\exp \left[ - \frac{y}{2} [(1 + c(y/L)^{1/\alpha})^{-\alpha} + (1 - c (y/L)^{1/\alpha})^{-\alpha}]\right] \to \frac{2^L}{L}
\end{equation}
as claimed.

\begin{figure*}[ht!]
    \begin{minipage}{.49\textwidth}
     \includegraphics[width=\textwidth]{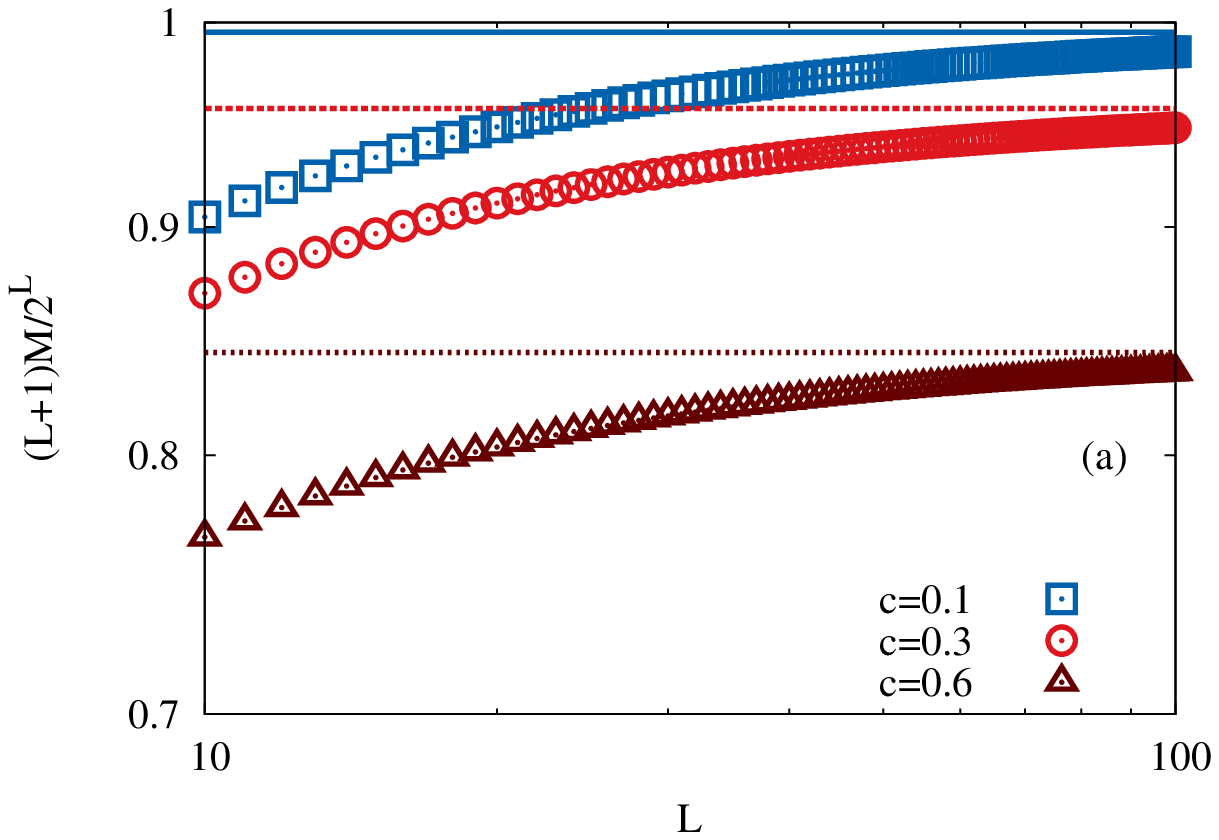}
    \end{minipage}
    \hspace{.5cm}
    \begin{minipage}{.49\textwidth}
     \includegraphics[width=\textwidth]{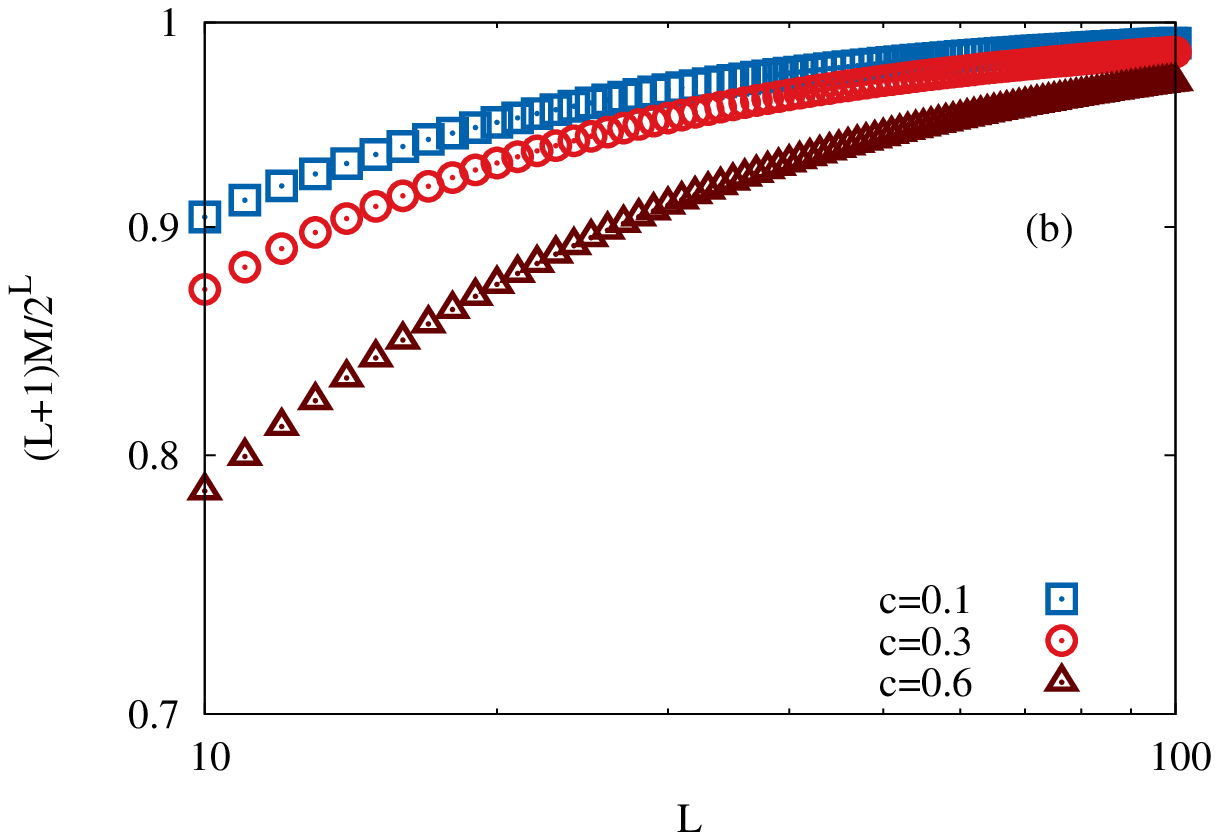}
    \end{minipage}
    \caption{Expected number of maxima $\cal{M}$ normalized by its asymptotic value for a HoC
landscape, $\frac{2^L}{L}$. Panel (a) shows the exact result
in Equation \ref{M:exp2} for an exponentially distributed random
contribution. Dashed horizontal lines illustrate that the ratio converges to
$\frac{1}{\cosh(c)}$ for large $L$. 
Panel (b) shows a numerical evaluation of Equation \ref{eq:nomsum} 
for a Pareto distributed random contribution with $\alpha = 2$ ($\kappa = \frac{1}{2}$). In accordance
with Equation \ref{M:Pareto2} the ratio converges to unity for large $L$.}
    \label{fig:compnom}
\end{figure*}

\subsubsection{Gumbel class.} We are now in the position to generalize the results obtained
for the exponential and Gumbel distributions to the entire Gumbel class of EVT. For calculational
convenience we choose the one-parameter family of Weibull distributions defined by 
\label{eq:genGumbel}
\begin{align}
     P(x) &=\begin{cases}
                            1-\exp[-x^\beta] & x \geq 0\\
                            0 & x < 0
                         \end{cases}
\end{align} 
to represent Gumbel-class distributions with tails that are heavier (for $\beta < 1$)
or lighter (for $\beta > 1$) than exponential. As in the cases considered above, the dominant
contribution to the number of maxima comes from the part of the
integral in Equation \ref{M:general} that extends from
$c$ to infinity, which now reads
\begin{equation}
\label{M:genGum}
\mathcal M \approx 2^L \int_c^\infty dx \; p(x) \left[ 1 - \frac{1}{2} 
[e^{-(x+c)^\beta} + e^{-(x-c)^\beta}] \right]^L.
\end{equation}
The crucial difference between the cases $\beta < 1$ and $\beta > 1$ lies in the behavior 
of the ratio of the two exponential terms inside the inner square brackets. For $\beta < 1$ 
\begin{equation}
\label{M:exps}
\lim_{x \to \infty} \frac{e^{-(x+c)^\beta}}{e^{-(x-c)^\beta}} = 1
\end{equation}
which implies that the shift by $\pm c$ becomes irrelevant asymptotically and 
${\cal{M}} \to \frac{2^L}{L}$ independent of $c$, as in the Fr\'echet class. On the other
hand, for $\beta > 1$ one finds that $e^{-(x-c)^\beta} \gg e^{-(x+c)^\beta}$ for large 
$x$ and Equation \ref{M:genGum} simplifies to 
\begin{equation}
\label{M:genGum2}
\mathcal M \approx 2^L \int_c^\infty dx \; p(x) \left[ 1 - \frac{1}{2}e^{-(x-c)^\beta} \right]^L
\approx 2^L \int_c^\infty dx \; p(x) \exp\left[-\frac{L}{2} e^{-(x-c)^\beta}\right].
\end{equation}
For large $L$ the integrand is effectively zero below a cutoff scale $x^\ast$ determined by
$L e^{-(x-c)^\beta} \sim 1$ or $x^\ast \approx c + (\ln L)^{1/\beta}$. Thus
\begin{equation}
\label{M:genGum3}
\mathcal M \approx 2^L [1-P(x^\ast)] = 2^L \exp[-(c+(\ln L)^{1/\beta})^\beta] \approx 
\frac{2^L}{L} \exp[-\beta c (\ln L)^{1 - \frac{1}{\beta}}]
\end{equation}
to leading order in $L$.
 
\subsubsection{Weibull class.} We next consider distributions with bounded
support, which we take to be the unit interval $[0,1]$. In the
evaluation of the general expression in Equation \ref{M:general} it has to be
taken into account that $P(x-c) = 0$ for $x < c$ and $P(x+c) = 1$ for
$x > 1-c$, which implies in particular that the cases $c <
\frac{1}{2}$ (where $c < 1-c$) and $c > \frac{1}{2}$ (where $c > 1-c$)
have to distinguished. We begin with the uniform distribution and
assume $c < \frac{1}{2}$, which yields
\begin{eqnarray}
\label{M:uniform1}
\mathcal M = \int_0^c dx \; (x+c)^L + \int_c^{1-c} dx \; (2x)^L +
\int_{1-c}^1 dx \; (1 + x - c)^L \nonumber \\
= \frac{1}{L+1}[(2-c)^{L+1} + (2^L-1) c^{L+1} - 2^L(1-c)^{L+1}].
\end{eqnarray} 
Analogously for $c > \frac{1}{2}$ we obtain
\begin{eqnarray}
\label{M:uniform2}
\mathcal M = \int_0^{1-c} dx \; (x+c)^L + \int_{1-c}^c dx \; +  
\int_{c}^1 dx \; (1 + x - c)^L \nonumber \\
= \frac{1}{L+1}[(2-c)^{L+1} -c^{L+1}] + 2c-1.
\end{eqnarray} 
Comparing Equations \ref{M:uniform1} and \ref{M:uniform2} the two expressions are seen to 
conincide at $c=\frac{1}{2}$. Moreover, Equation \ref{M:uniform1} reduces to $\frac{2^L}{L+1}$
for $c=0$ and Equation \ref{M:uniform2} confirms that ${\cal{M}} = 1$ for $c=1$, as expected.

As representatives of Weibull-class distributions with a general EVT index $\kappa < 0$ we consider
the Kumaraswamy distributions defined by 
\label{eq:genWeibull}
\begin{align}
     P(x) &=\begin{cases}
                            1-(1-x)^\nu & 0 \leq x \leq 1\\
                            0 & x < 0 \\
			    1 & x > 1	
                         \end{cases}
\end{align} 
with $\nu = - \frac{1}{\kappa}$. We focus on the leading order behavior of the number of maxima
for large $L$, which is given by the part of the integral in Equation \ref{M:general} that contains
the upper boundary of the support at $x=1$ (compare to Equations \ref{M:uniform1} and \ref{M:uniform2}). 
We assume $c < \frac{1}{2}$ and obtain
\begin{equation}
\label{M:genWeib1}
\mathcal M \approx 2^L \int_{1-c}^1 dx \; \nu (1-x)^{\nu-1} \left[ 1 - \frac{1}{2} (1 - x + c)^\nu 
\right]^L = 2^L \int_0^c dy \; \nu y^{\nu-1} \left[ 1 - \frac{1}{2} (y+c)^\nu \right]^L,
\end{equation} 
which is dominated by the region near $y=0$ for large $L$. Expanding $(y+c)^\nu$ for small $y$
it follows that
\begin{equation}
\label{M:genWeib2}
\mathcal M \approx (2-c^\nu)^L \int_0^c dy \; \nu y^{\nu-1} \exp 
\left[-\frac{\nu c^{\nu-1} L y}{2-c^\nu} \right] \to \frac{\nu \Gamma(\nu)}{(\nu c^{\nu-1})^\nu}
\frac{(2-c^\nu)^{L+\nu}}{L^\nu}
\end{equation}
for large $L$, where $\Gamma$ denotes the Gamma-function. The result for $c > \frac{1}{2}$ is 
asymptotically the same. Thus to leading order the number of 
maxima is ${\cal{M}} \sim \frac{(2-c^{-\frac{1}{\kappa}})^L}{L^{-\frac{1}{\kappa}}}$ in the 
Weibull class.
 
\subsubsection{Expansion for small $c$.} A unified view of the effect of the fitness gradient
on the expected number of maxima can be obtained by expanding the general expression 
in Equation \ref{M:general} for small values of $c$.
To leading order in $c$ the integrand is $P(x-c) + P(x+c) = 2 P(x) + c^2 P''(x) = 2  P(x) + c^2 p'(x)$,
where primes indicate derivatives with respect to $x$. Integrating and keeping terms of order $c^2$ we thus obtain
\begin{equation}
\label{Mexpand}
{\cal{M}} = \frac{2^L}{L+1} - 2^{L-2}L(L-1)c^2 \int dx \; p(x)^3 P(x)^{L-2} + 
{\cal{O}}(c^4).
\end{equation}
Evaluating the integral on the right hand side for the GPD
distribution in Equation \ref{GPD} one finds
that it is of the order of $L^{-(3+2\kappa)}$, and hence the entire correction term is
of order $2^L L^{-(1+2 \kappa)}c^2$. Thus for $\kappa > 0$ (Fr\'echet class) the correction 
becomes negligible compared to the HoC term for large $L$.
In contrast, for $\kappa < 0$ (Weibull class) the correction term eventually dominates 
the HoC term, showing that the leading order behavior of ${\cal{M}}$ is modified. 

\subsection{Location of the global maximum.}
To compute the probability $\tilde{P}_\mathrm{max}(d)$ that a sequence
at distance $d$ is the \textit{global} fitness maximum one has to
demand that its fitness exceeds the fitnesses of all other sequences,
which leads to the expression 
\begin{equation}
 \tilde{P}_\mathrm{max}(d)=\int_{-\infty}^{\infty}\mathrm{d}x p(x+cd)\prod_{j=0}^{L}P(x+cj)^{\binom{L}{j}-\delta_{dj}}.\label{bla}
\end{equation}
Inserting the Gumbel distribution in Equation \ref{eq:Gumbel} and using its shift
property in Equation \ref{eq:Pshift}, this can be evaluated according to 
\begin{eqnarray}
\label{global_dist}
 \tilde{P}_\mathrm{max}(d) &=& \int dy \; p_G(y) \prod_{j=0}^{L}P_G(y+c(j-d))^{\binom{L}{j}-\delta_{dj}} = \int
 dy \; p_G(y) P_G(y)^{\sum_{j=0}^L e^{-c(j-d)}\left[\binom{L}{j}
     - \delta_{dj}\right]} \nonumber\\
&=& \int dy \frac{dP_G}{dy} P_G(y)^{e^{cd}(1+e^{-c})^L-1} 
= \int_0^1 dz \; z^{e^{cd}(1 + e^{-c})^L - 1} =
\frac{e^{-cd}}{(1 + e^{-c})^L}.
\end{eqnarray}
For general fitness distributions one has to resort to an expansion in
$c$. Starting from Equation \ref{bla} and collecting terms linear in $c$ one
obtains 
\begin{equation}
\label{Ptildelinear}
\tilde{P}_\mathrm{max}(d) = \frac{1}{2^L}+c2^L(L/2-d)I_{2^L-2} +
\mathcal O(c^2),
\end{equation}
where 
\begin{equation}
\label{IL}
I_{L} = \int dx \; p(x)^2 P(x)^{L}.
\end{equation}
Expressions for $I_L$ for representatives 
of the three extreme value classes have been derived by
\citeN{jaspergregor}. For large $L$ they behave as \cite{Wergen2011}
\begin{equation}
\label{I_L} 
I_L \sim L^{-(2+\kappa)}. 
\end{equation}
 The weighted average with respect to $d$ then yields the approximate
expression 
\begin{equation}
 \mathbb{E}(d) \approx \frac{L}{2}-cL2^{2L-2} I_{2^L-2}
\end{equation}
for the mean distance of the global maximum to the reference
sequence. Using the asymptotic behavior of $I_L$ given in Equation \ref{I_L}
it follows that the shift in the position of the maximum from the HoC
value $L/2$ is of order $c L 2^{-\kappa L}$.

\renewcommand{\theequation}{B\arabic{equation}}
\setcounter{equation}{0}
\renewcommand{\thefigure}{B\arabic{figure}}
\setcounter{figure}{0}
\section{Appendix B: Calculation of the fitness correlation function}
\label{app:B}

For notational convenience we substract the mean value
$\mathbb{E}(\eta)$ of the random fitness component from Equation \ref{RMF}
and define
\begin{equation}
\label{eq:f}
f(\sigma) = F(\sigma) - \mathbb{E}(\eta) = - c d + \xi(\sigma) \;\;\;
\text{with} \;\;\; \xi(\sigma) = \eta(\sigma) -  \mathbb{E}(\eta),
\end{equation}
such that $\mathbb{E}(\xi) = 0$. It is then easy to see that
\begin{equation}
\label{eq:fcorr}
\langle(F(\sigma)-\langle
   F(\sigma)\rangle)(F(\sigma')-\langle F(\sigma') \rangle)\rangle_r = \langle(f(\sigma)-\langle
   f(\sigma)\rangle)(f(\sigma')-\langle f(\sigma') \rangle)\rangle_r =
   c^2 [\langle d d' \rangle_r - \langle d \rangle^2] + v
   \delta_{\sigma,\sigma'}
\end{equation}
where $v = \mathrm{Var}(\eta)$ is the variance of $\eta$ (or $\xi$), which is assumed in the 
following to exist, and $d = D(\sigma,\sigma^\ast)$,  $d' =
D(\sigma',\sigma^\ast)$. While the sequence space averages of $d$ and
$d'$ are obviously equal to $\frac{L}{2}$, the evaluation of $\langle
d d' \rangle_r$ requires a double summation, first over sequences $\sigma$ 
at distance $d$ from $\sigma^\ast$ and then over sequences $\sigma'$
at distance $r$ from $\sigma$. The latter sequences are 
grouped according to a number $k$ that counts how many of the $r$ point mutations 
distinguishing $\sigma^\prime$ from $\sigma$ fall on alleles that are
different in $\sigma$ and $\sigma^\ast$. 
Obviously, each such mutation decreases the distance $d'$ by 1, while each of the $r-k$ mutations 
acting on previously unaltered alleles increases $d'$ by 1, such that 
$d' = d - k + (r-k) = d + r - 2k$. The number of sequences $\sigma'$
with a given value of $k$ is equal to $\binom{d}{k} \binom{L-d}{r-k}$.
Thus the sum to be evaluated is 
\begin{align}
\langle d d' \rangle_r&=\frac{1}{2^L}\sum_{d=0}^L
\binom{L}{d}\frac{d}{\binom{L}{r}}\sum_{k=0}^r
\binom{L-d}{r-k}\binom{d}{k}(d+r-2k) \nonumber\\
&=\frac{1}{2^L}\sum_{d=0}^L
\binom{L}{d}d\left[d\left(1-\frac{2r}{L}\right)+r\right] \nonumber\\
&=\frac{L^2}{4}
+ \frac{L}{4}\left(1 - \frac{2r}{L}\right) = \langle d \rangle^2 + \frac{L}{4}\left(1 - \frac{2r}{L}\right),
\end{align}
where the combinatorial identities 
$$
\sum_{k \geq 0} \binom{j}{l+k} \binom{m}{n-k} = \binom{j+m}{l+n}
\;\;\; \text{and} \;\;\; k \binom{n}{m} = n \binom{n-1}{k-1}
$$
have been used \cite{concrete_math}. Putting everything together finally yields 
Equation \ref{eq:CRMF}. 

\renewcommand{\theequation}{C\arabic{equation}}
\setcounter{equation}{0}
\renewcommand{\thefigure}{C\arabic{figure}}
\setcounter{figure}{0}
\section{Appendix C: Calculation of the number of exceedances}
\label{app:C}
We have seen above in MODEL that the full neighbourhood $\nu(\sigma)$ of a sequence $\sigma$ with $D(\sigma,\sigma_0)=d$
    is divided into the uphill neighbourhood 
    with the corresponding distribution function
    $P^\uparrow(x)= P(x+c(d-1))$ of fitness values,
    $\sigma$ itself with distribution function $P^\bullet(x) = P(x+cd)$, and the downhill neighbourhood 
    with distribution function
    $P^\downarrow(x) = P(x+c(d+1))$.
    The full distribution function of fitness values is then given by
    \begin{align}
\label{Pidist}
        \Pi(x) =&\frac{1}{L+1}\left(d P^\uparrow(x)+P^\bullet(x)+(L-d) P^\downarrow(x)\right)\\
        &= \frac{1}{L+1}\left(d P(x+c(d-1)) + P(x+cd)+(L-d) P(x+c(d+1))\right),\nonumber
    \end{align}
and the expectation of the $k$th largest fitness value is obtained as \cite{orderstat}
        \begin{equation}
\label{muk}
            \mu_k = (L+1) \binom{L}{L+1-k}\int_0^1 x\Pi(x)^{L-k+1}(1-\Pi(x))^{k-1}d\Pi(x).
        \end{equation}
In general, the evaluation of this expression is complicated, because
the different components of $\Pi$ do not have the same support. Here
we show how this problem can be circumvented in an approximate way for
the special case of an exponential distribution fitness distribution
$P(x) = 1 - e^{-x}$. Naively inserting this into Equation \ref{Pidist} 
we obtain   
    \begin{equation}
\label{Pshifted}
        \Pi(x)  = 1-e^{-x+\log\left(\frac{1}{L+1}e^{-cd}\left(d
              e^{c}+1+(L-d)e^{-c})\right)\right)} =: 1-e^{-x+\log(\xi(c,d,L))}=P(x-\log(\xi(c,d,L)))
    \end{equation}
with 
\begin{equation}
\label{xidef}
\xi(c,d,L) = \frac{e^{-cd}}{L+1}\left(de^c + 1 + (L-d)e^{-c} \right).
\end{equation}
Our approximation consists in ignoring the fact that
Equation \ref{Pshifted} only holds on the intersection of the supports of
$P^\uparrow$, $P^\bullet$ and $P^\downarrow$, and instead defining the  
\textit{common support} of $\Pi(x)$ by $[\log(\xi(c,d,L)),\infty)$,
such that $\Pi(\log(\xi)) = 0$. Thus the full distribution of fitness
values defined in Equation \ref{Pidist} is replaced by a simple exponential that
is shifted in a $d$-dependent way. 

To proceed we recall that the expected value $m_{k,n}$ of the $k$th largest
out of $n$ exponentially distributed random variables is given by
\cite{orderstat}
\begin{equation}
\label{mkn}
m_{n,k} = H_n - H_{k-1} \approx \log\left(\frac{n}{k-1}\right) 
\end{equation}
where $H_n=\sum_{k=1}^{n}\frac 1 k$ are the harmonic numbers and we use the convention
that $H_0 = 0$. In the second part of Equation \ref{mkn}  the logarithmic
approximation $H_n \approx \log(n) + \gamma$ valid for large arguments
has been applied, with $\gamma \approx 0.5772156649...$. It follows that the
mean of the $k$th largest fitness value in the neighborhood, defined
in Equation \ref{muk}, is approximately given by  
    \begin{equation}
\label{mukexp}
        \mu_k =\log(\xi(c,d,L))+H_{L+1}-H_{k-1}  \approx \log\left(\frac{e^{-cd}}{k-1}\left(de^c+(L-d)e^{-c}+1\right)\right).
    \end{equation}
     
\subsection{After a step up.}
   To obtain the mean number of exceedances (NoE) after a transition from a sequence $\sigma$ at distance $d$ to
    $\sigma'$ at
    distance $d-1$, where $\sigma'$ has rank $r$ in the old
    neighborhood, we need to count how many times $F(\sigma')$ 
is exceeded in the new neighborhood. As an estimation, $F(\sigma') \approx \mu_r(L,c,d)$ is compared to
    the mean rank-ordered fitness values from the uphill and the
    downhill parts of the new neighborhood, which are sets of $d-2$
    and $L-d+1$ independent exponential variables, respectively. 
In the uphill neighborhood at distance $d-2$ from reference sequence the exponential random variables are
shifted by $c(d-2)$, and therefore the number of exceedances $k_>$
derived from this part of the neighborhood is obtained by solving the
relation 
\begin{equation}
\label{kgt1}
\mu_r = m_{d-1,k_>} - c(d-2) \approx
\log\left(\frac{d-1}{k_>-1}\right) - c(d-2)
\end{equation}
for $k_>$. Using the approximation in Equation \ref{mukexp} this yields the
expression
\begin{equation}
\label{kgt2}
k_>^\mathrm{up} = 1 + \frac{(d-1)(r-1) e^{2c}}{d e^{c} + 1 +
(L-d) e^{-c}}.
\end{equation}
Similarly, the contribution $k_<$ to the exceedances from the downhill
neighborhood is obtained from the relation $\mu_r = m_{L-d+1,k_<} -
cd$, which yields
\begin{equation}
\label{ksm}
k_<^\mathrm{up} = 1 + \frac{(L-d+1)(r-1)}{de^c + 1 + (L-d) e^{-c}}.
\end{equation}
To complete the calculation we have to take into account the fact that, by construction,
$k_>^\mathrm{up} \leq d-1$ and $k_<^\mathrm{up} \leq L-d+1$, which is
not always satisfied by the approximate expressions in Equations \ref{kgt2} and
\ref{ksm}. Incorporating these constraints we arrive at our final result  
\begin{equation}
\label{Nup}
{\cal{N}}^\mathrm{up} = \min(k_>^\mathrm{up},d-1) +
\min(k_<^\mathrm{up},L-d+1).
\end{equation}
A simpler and more transparent expression can be obtained by assuming
that $d \gg 1$ and $e^c$ is not too large. Under the first assumption
the combination of Equations \ref{kgt2} and \ref{ksm} reduces to 
$k_>^\mathrm{up} + k_<^\mathrm{up} = 2 + (r-1)e^c$, while the second
assumption ensures that the $\min$-constraints in Equation \ref{Nup} can
be ignored, such that ${\cal{N}}^\mathrm{up} = k_>^\mathrm{up} +
k_<^\mathrm{up} = 2 + (r-1)e^c$, see Equation \ref{NoE_simple} in the main text.  

\subsection{After a Step Down.}
    \label{app:stepdown}
    The calculation of the NoE after a step is taken in the downhill
    direction is analogous to the previous one.
In this case the contribution from the upper part of the new
neighborhood is obtained from the relation $\nu_r = m_{d+1,k_>}-cd$,
which yields
\begin{equation}
\label{kgtdown}
k_>^\mathrm{down} = 1 + \frac{(r-1)(d+1)}{d e^c + 1 + (L-d) e^{-c}}.
\end{equation}
Correspondingly,
the contribution from the downhill part is obtained from evaluating
$\mu_r = m_{L-d-1,k_<} - c (d+2)$, with the result
\begin{equation}
\label{ksmdown}
k_<^\mathrm{down} = 1 + \frac{(r-1)(L-d-1)e^{-2c}}{d e^c + 1 +
  (L-d)e^{-c}}.
\end{equation}
The final estimate for the number of exceedances reads
\begin{equation}
\label{Ndown}
{\cal{N}}^\mathrm{down} = \min(k_>^\mathrm{down},d+1) +
\min(k_<^\mathrm{down},L-d-1),
\end{equation}
and the simplified expression in Equation \ref{NoE_simple} arises from the
same approximations employed previously for ${\cal{N}}^\mathrm{up}$. 

\begin{figure*}
    \begin{minipage}{.49\textwidth}
    \includegraphics[width=\textwidth]{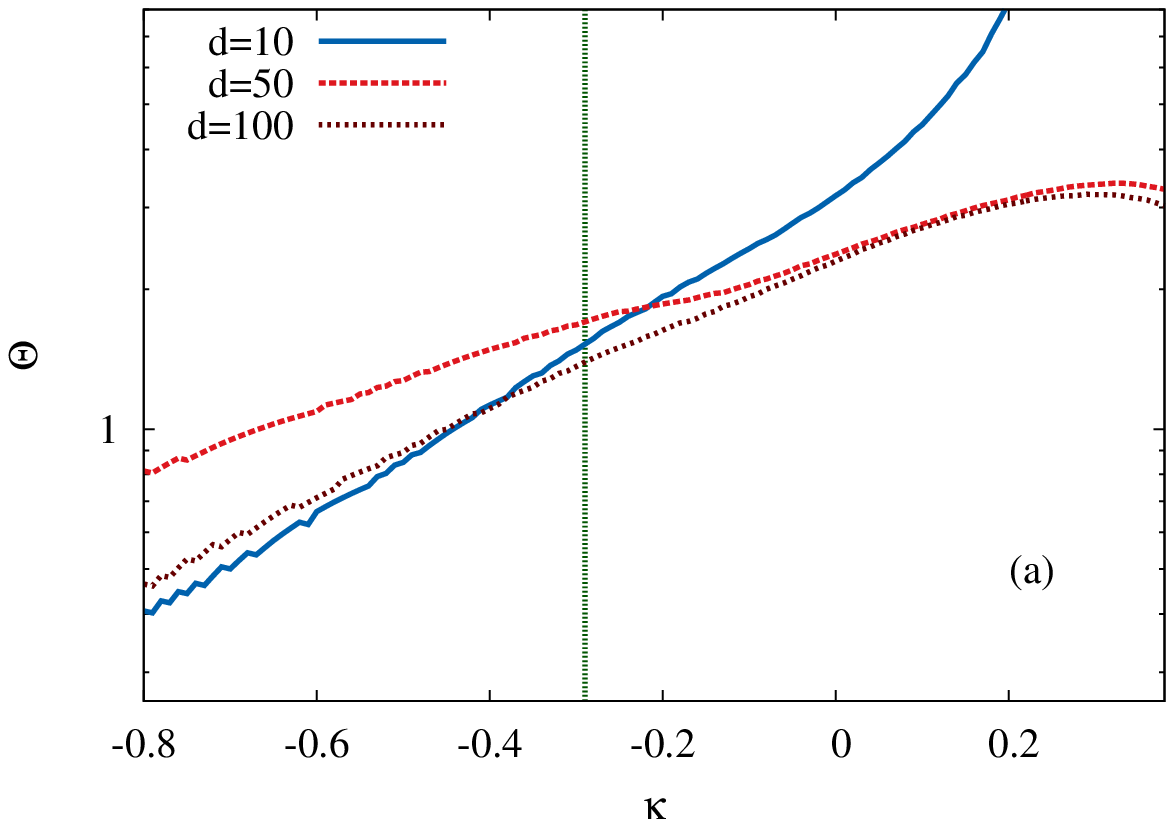}
    \end{minipage}
    \hspace{.5cm}
    \begin{minipage}{.49\textwidth}
    \includegraphics[width=\textwidth]{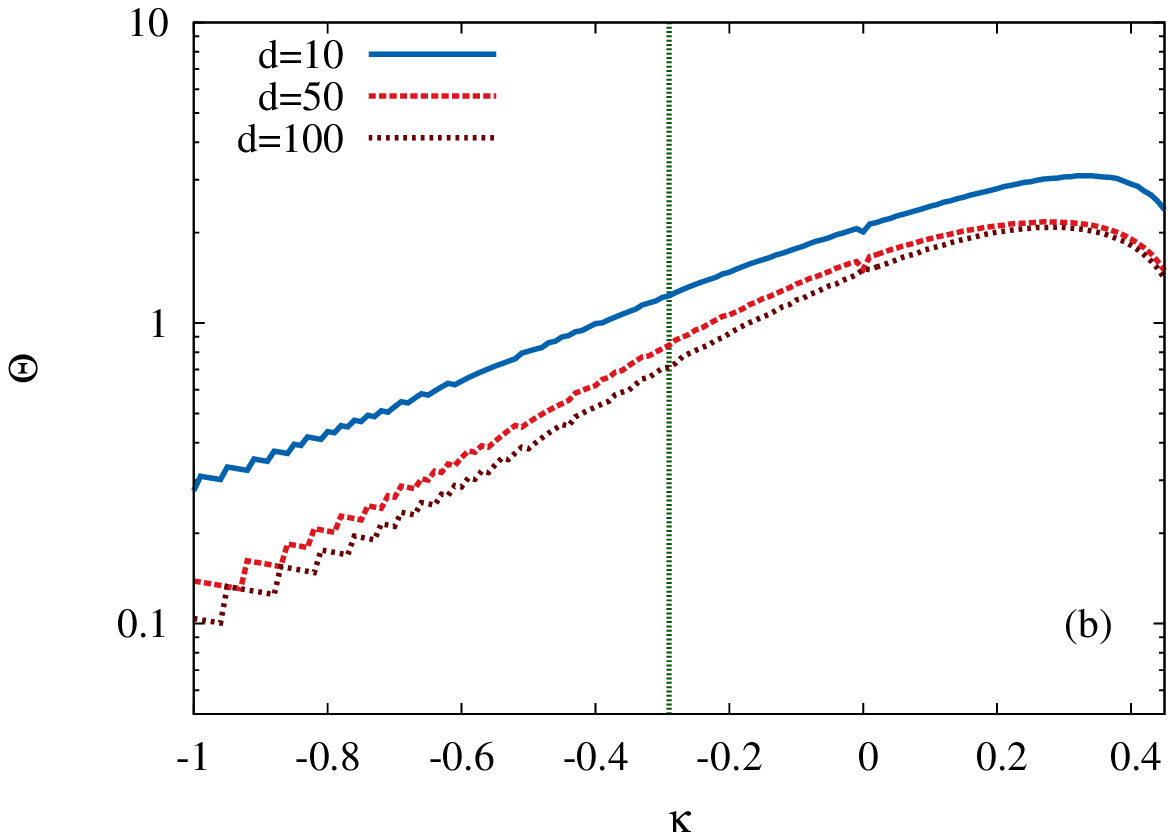}
    \end{minipage}
    \caption{The figure shows the minimal value of the scaled RMF parameter $\theta = \frac{c}{\sqrt{\mathrm{Var(\eta)}}}$
  required to generate on average 9 exceedances after an adaptive
  step. Panel
  (a) shows results for initial rank 1 and panel (b) for initial rank 3.
  The random
  fitness component is assumed to be distributed according to the GPD
  distribution with EVT index $\kappa$, and different curves
  correspond to different values of the initial distance $d$ to the
  reference sequence. The experimental estimate $\kappa \approx -0.29$
  of the EVT index is indicated by a vertical line.} 
  \label{fig:csigmexc}
\end{figure*}

\subsection{Analysis of the experiment of Miller et al. (2011).} In Figure 
\ref{fig:expsupp} we reported the RMF model parameter combinations $(\kappa, c)$ required to explain
the 9 beneficial second step mutations observed on the background of a highly fit first step mutation
in the experiment of \citeN{miller}. Figure \ref{fig:csigmexc} displays the same data with the additive
selection coefficient $c$ scaled by the standard deviation of the random fitness component, which in 
the case of the GPD distribution is given by 
\begin{equation}
\label{GPD_StD}
\sqrt{\mathrm{Var}(\eta)} = \sqrt{\frac{1}{(1-\kappa)^2(1-2\kappa)}}.
\end{equation}

\end{document}